# Sea Level and Socioeconomic Uncertainty Drives High-End Coastal Adaptation Costs

T. E. Wong[1], C. Ledna[2], L. Rennels[2], H. Sheets[1], F. C. Errickson[3], D. Diaz[4], and D. Anthoff[2]

[1] School of Mathematical Sciences, Rochester Institute of Technology, Rochester, New York USA.

[2] Energy and Resources Group, University of California Berkeley, Berkeley, CA, USA.

[3] School of Public and International Affairs, Princeton University, Princeton, NJ, USA.

[4] Energy Systems and Climate Analysis Group, Electric Power Research Institute, Palo Alto, CA, USA.

Corresponding author: Tony Wong (aewsma@rit.edu)

**Key Points:**

- The common practice to use just a few percentiles to represent sea-level uncertainties can dramatically underestimate high-end coastal risks
- Socioeconomic uncertainties, especially the value of at-risk lands, are the dominant drivers of uncertainty surrounding adaptation costs
- Coastal adaptation decisions weigh up-front adaptation costs against uncertain future losses of property and population

**Abstract**

Sea-level rise and associated flood hazards pose severe risks to the millions of people globally living in coastal zones. Models representing coastal adaptation and impacts are important tools to inform the design of strategies to manage these risks. Representing the often deep uncertainties influencing these risks poses nontrivial challenges. A common uncertainty characterization approach is to use a few benchmark cases to represent the range and relative probabilities of the set of possible outcomes. This has been done in coastal adaptation studies, for example, by using low, moderate, and high percentiles of an input of interest, like sea-level changes. A key consideration is how this simplified characterization of uncertainty influences the distributions of estimated coastal impacts. Here, we show that using only a few benchmark percentiles to represent uncertainty in future sea-level change can lead to overconfident projections and underestimate high-end risks as compared to using full ensembles for sea-level change and socioeconomic parametric uncertainties. When uncertainty in future sea level is characterized by low, moderate, and high percentiles of global mean sea-level rise, estimates of high-end (95th percentile) damages are underestimated by between 18% (SSP1-2.6) and 46% (SSP5-8.5). Additionally, using the 5th and 95th percentiles of sea-level scenarios underestimates the 5-95% width of the distribution of adaptation costs by a factor ranging from about two to four, depending on SSP-RCP pathway. The resulting underestimation of the uncertainty range in adaptation costs can bias adaptation and mitigation decision-making.

**Plain Language Summary**

Rising global sea levels and intensifying storms cause risks for people and properties in low-lying coastal areas. Strategies to manage these risks include protective measures like building seawalls, elevating existing structures, and relocation away from the coast. Here, we use a computer model to examine how coastal areas can use these adaptation strategies to mitigate damages from sea-level rise and coastal flooding. We find that characterizing uncertainty by using only representative low, moderate, and high sea-level rise cases, high-end total global adaptation costs are underestimated by 18-46% relative to an ensemble modeling approach that more carefully samples future sea-level rise and socioeconomic uncertainties. Even in a high greenhouse gas emissions scenario, investments in relocation and coastal defenses can reduce global annual losses of property and lives from flooding by over US$1 trillion in the year 2100. These results highlight the social and economic need for climate mitigation and adaptation to manage coastal risks around the world.

**1 Introduction**

Sea-level rise poses risks both to coastal areas and the people who inhabit them around the world. Recently, the Intergovernmental Panel on Climate Change's Sixth Assessment Report (AR6) found with high confidence that global mean sea levels will rise between 0.1 and 0.6 m by 2050 (relative to 1995-2014 mean sea level), depending on the emissions scenario (Fox-Kemper et al., 2021). By the year 2100, these projections are estimated to reach up to 1.0 m (95th percentile) in the low-/moderate-emissions Representative Concentration Pathway 2.6 and 4.5 scenarios (RCP2.6 and 4.5). In the high-emissions RCP8.5 scenario, these projections can reach up to 1.6 or 2.4 m (95th percentile), depending on the modeling assumptions regarding the major ice sheets. Recent work has also made advances to estimate the number of people globally who live in at-risk coastal areas. Kulp & Strauss (2019), for example, estimate that just over one billion people live in low elevation coastal zones. Of those billion people, about 37 million live in the United States, while close to 680 million live in just eight Asian countries. Through improvements to digital elevation maps, those authors also find an estimated 110 million people globally living at or below current high tide levels (Kulp & Strauss, 2019).

In order to mitigate future losses from damage and displacement to property and people, it is necessary to manage the risks posed by rising sea levels. This task demands information that can be provided by combining two types of models: i) models to connect deeply uncertain future greenhouse gas and climate change scenarios to projections of global and local mean sea-level rise, and ii) models to link those scenarios of local mean sea-level rise to coastal adaptation decision-making. When generating projections of future sea-level rise for informing adaptation decision-making, there are obvious benefits to employing highly detailed physical and socioeconomic models. However, given a fixed overall computational budget (in terms of numbers of machines and/or wallclock hours), this level of detail must be balanced against the efficiency of those models, and the computational demands of the statistical models to be employed for characterizing uncertainty (Wong, Bakker, Ruckert, et al., 2017).

On one hand, detailed global climate models/general circulation models (GCMs) offer an approach for physical modeling of the many interrelated components of the climate system (for example, the Coupled Model Intercomparison Project 6 (Eyring et al., 2016)). However, recent work has found that downscaling of twenty-first century GCM ocean dynamic sea-level changes over the Northwestern European Shelf region can be significantly biased by unresolved processes in the GCM (Hermans et al., 2020). Representations of ice sheet dynamics within GCMs, in particular, are often simpler than in regional climate models – or potentially even absent - but can better account for the many feedbacks between the ice sheets, overlying atmosphere, and the surrounding ocean (Alexander et al., 2019). These potential drawbacks, as well as the high computational costs associated with their large spatial and/or temporal domains, point to the importance of balancing these detailed physical models with similarly sophisticated statistical modeling methods. As state-of-the-art Bayesian model calibration methods are also often computationally expensive, the high computational expense associated with GCMs generally precludes the use of Bayesian methods for their calibration, with exceptions for small spatial or temporal scales (e.g., Wong, Kleiber, & Noone, 2017). On the other hand, semi-empirical models (SEMs) and model emulation offer computationally efficient alternatives to GCMs (Kopp et al., 2014; Mengel et al., 2016; Nauels et al., 2017; Rahmstorf, 2007; Wong, Bakker, Ruckert, et al., 2017). SEMs can include through stylized parameterizations an accounting of processes that GCMs either represent poorly or do not represent at all.

The computational efficiency of this stylization also enables the generation of full probability distributions for events of interest, which in turn provides estimates of the tails of these distributions. This is important for processes such as precipitation or local sea level, where the upper tails of these events' probability distributions represent low-probability but high-consequence outcomes. Specifically, for flood risk, extreme value distributions govern the occurrence and frequency of extreme coastal sea levels (Coles, 2001). Resolving the tails of these distributions is critical for characterizing coastal hazards driven by extreme sea levels (Wahl et al., 2017). Estimating these distributions via model simulation is a common practice (e.g., Buchanan et al., 2017; Lee et al., 2017; Rashid et al., 2019; Ruckert et al., 2019; Wong, Sheets, et al., 2022), but often requires large ensemble sizes. This demand for large ensembles places strict importance on computational efficiency. SEMs thus also offer a viable computationally efficient avenue to provide physical model projections that can usefully inform decision analyses for adaptation to extreme coastal sea levels.

Given a set of sea-level projections, coastal managers can decide to make different adaptation decisions to protect their local communities. These can include *protection* via "hard" infrastructure such as seawalls, dikes, and bulkheads as well as "soft" measures such as beach nourishment; *accommodation* via measures such as warning systems and raising existing infrastructure; *advance* by infilling and protecting once-inundated land; *retreat*, via relocation of people and assets; and *ecosystem-based adaptation*, such as the maintenance of wetlands and reefs (Oppenheimer et al., 2019). For simplicity's sake, we will refer generally to all hard forms of protection as "seawalls". Protection can also be interpreted more generally to include accommodation measures such as elevating existing structures, although cost estimates will of course not represent these other adaptation options. Retreat refers to moving people and property further inland to avoid the encroaching seas. Retreat can be either proactive (as a preventative, managed adaptation decision) or reactive (in response to a location becoming inundated and unlivable).

Within a particular choice of adaptation action, decision-makers can also implement different levels of that adaptation. For example, New Orleans, Louisiana is mandated to generally protect against the 100-year flood hazard level (that is, a storm whose severity is expected to be seen once per 100 years, on average) (Coastal Protection and Restoration Authority of Louisiana, 2017). This can be achieved, for example, via seawall protection, or alternatively through relocation sufficiently far inland that the same level of protection is offered, in terms of reduced risk of flooding. Retreat may not be an attractive adaptation option for New Orleans because of the high costs of relocating a major metropolitan area and trade hub. This is represented in models to inform adaptation decision-making through some decision criterion, such as minimizing the expected total cost over a decision-relevant time horizon. For coastal adaptation models, this expected total cost can include the cost of constructing new flood defenses, the cost of relocating and reconstructing infrastructure further inland, and damages associated with flooding, inundation, and wetland loss (e.g., Diaz, 2016; Lincke & Hinkel, 2018; Tiggeloven et al., 2020).

One such coastal adaptation model is the Coastal Impact and Adaptation Model (CIAM) (Diaz, 2016). CIAM determines the costs and damages associated with different modes of adaptation to future coastal hazards. Owing to the limitations of data to constrain cost estimates for the many possible forms of coastal adaptation, within CIAM, coastal areas can take no action (Figure 1b), or can implement adaptation in the form of retreat from the coast (Figure 1c) or protection via construction of seawalls (Figure 1d). This constraint is characteristic of other contemporary large-scale studies in coastal adaptation (e.g., Abadie et al., 2019; Hinkel et al., 2014; Jevrejeva et al., 2018; Lincke & Hinkel, 2018; Prahl et al., 2018; Rohmer et al., 2021; Tiggeloven et al., 2020; Vousdoukas et al., 2020). Current research is limited in its ability to examine this full suite of adaptation measures in a global-scale cost-benefit analysis due to a lack of comprehensive global data, and most of the aforementioned studies restrict their focus to protection via seawall construction and maintenance. Thus, the ability of CIAM to examine the degree to which both protection *and* relocation play important roles in adapting coastal areas is an important feature of the model. However, at present, CIAM does not represent potential negative impacts of planned relocation or seawall protection projects, such as loss of coastal habitats or agricultural productivity (Portner et al., 2022).

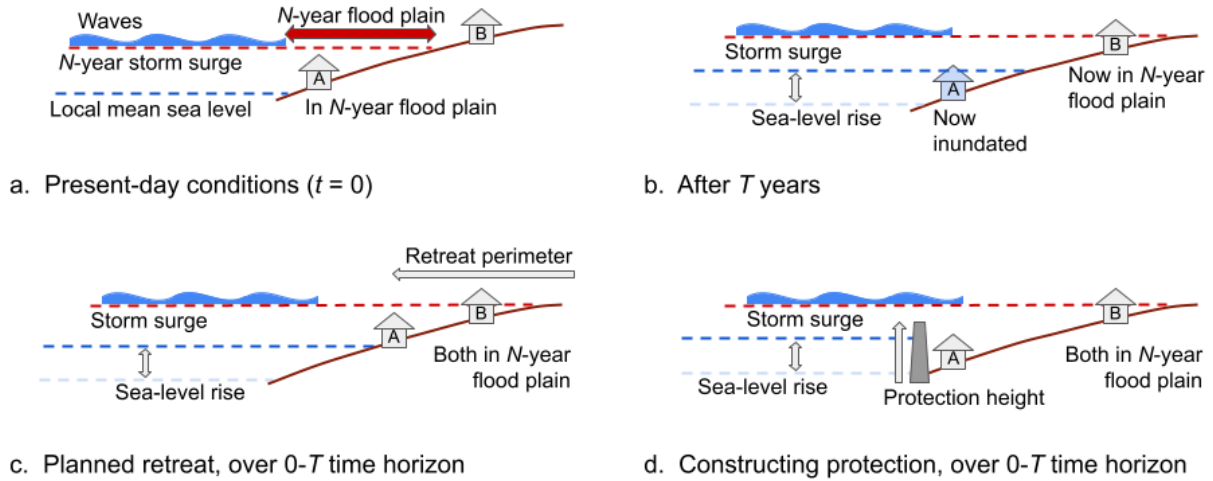

**Figure 1.** Adaptation options within CIAM and the total coastal hazard. a) Conditions at the beginning of an adaptation period in which structure A is in the *N*-year floodplain and structure B is not. b) After *T* years, at the end of the adaptation period, local mean sea-level rise has inundated the area occupied by structure A, and structure B is now in the *N*-year floodplain. c) At the end of the adaptation period, with planned retreat, both structures are in the *N*-year floodplain but not inundated. d) At the end of the adaptation period, with the construction of seawall protection, neither structure is on inundated land, but both are in the *N*-year floodplain.

The adaptation decisions each have different costs for different sections of the coast. CIAM considers adaptation actions and costs for the 12,148 coastal segments of the Database for Impact and Vulnerability Analysis (DIVA) database (Vafeidis et al., 2008). CIAM considers a range of adaptation levels for each of retreat and construction of protection, as well as no adaptation. For each of the considered combinations of adaptation strategy and level, CIAM computes the decision that minimizes the net present value of expected cost over a specified time horizon. The design of CIAM and other coastal adaptation models requires estimates of future sea levels, which provide the future conditions against which local coastal adaptation decisions will be made. Previous work using CIAM has assumed that future sea-level rise is known perfectly over the entire time horizon for the model simulation (2010-2200; Diaz, 2016). However, there is only strong agreement among global mean sea-level projections until about 2050 (Fox-Kemper et al., 2021). After that, future sea-level changes and associated coastal hazards are deeply uncertain, owing largely to uncertainties related to the future of the Antarctic ice sheet (Hough & Wong, 2022; Kopp et al., 2017; Le Bars et al., 2017; Wong, Bakker, & Keller, 2017). This limited information can have substantial consequences in terms of increased costs (Diaz & Keller, 2016; Hall et al., 2012). In light of these uncertainties, it is necessary to examine how limited foresight of future sea levels affects decision-makers' ability to manage coastal risks, and to use projections of sea-level changes accounting for these deep uncertainties in ice sheet dynamics.

The DIVA database is well-used in a variety of studies of coastal risks and adaptation costs. Hinkel et al. (2014) apply DIVA to explore impacts of increased coastal flooding on population and assets on a global scale, and examine how seawall construction/augmentation can mitigate the damages from this flooding. That work finds that by 2100, expected annual losses ranged from 0.3-5.0% of global gross domestic product (GDP) in their counterfactual constant-protection scenario under RCP2.6, and 1.2-9.3% of global GDP under RCP8.5 in this same constant-protection scenario. In another study using the DIVA database and SSP scenarios, Jevrejeva et al. (2018) find that the sea-level rise associated with unrestricted warming above 2 °C can lead to damages of nearly 3% of global GDP by 2100, without further coastal adaptation. However, proactive relocation and changing the protection standards for a given area are not considered in that work. Rather, that study considers a no-adaptation future and a business-as-usual adaptation scenario, and focuses on how temperature targets affect coastal impacts. To further explore and attribute coastal impacts to geophysical drivers, Rohmer et al. (2021) use random forests to decompose uncertainty in expected annual damages and adaptation costs into portions attributable to different geophysical and socioeconomic factors. That work finds that adaptation costs are largely driven by RCP scenario, and uncertainty in human activities (both SSP and RCP pathways) are the largest driving factor in future flood risks. Lincke & Hinkel (2018) tackle another key uncertainty using DIVA to examine the spatial heterogeneity in economically-efficient adaptation response to rising coastal sea levels, finding that it is generally economically efficient to enhance protection of urban areas, and to retreat from rural coastal areas.

Large-scale studies on coastal adaptation typically consider a counterfactual no-adaptation scenario as a baseline (e.g., Abadie et al., 2020; Diaz, 2016; Hinkel et al., 2014; Jevrejeva et al., 2018; Tiggeloven et al., 2020; Vousdoukas et al., 2020) and a scenario in which coastal communities pursue either economically-efficient (generally minimizing expected costs) adaptation (Tiggeloven et al., 2020; Vousdoukas et al., 2020) or augmenting defenses to maintain current levels of protection (e.g., Hinkel et al., 2014; Jevrejeva et al., 2018; Tiggeloven et al., 2020). The work of Tiggeloven et al. (2020) further considers adaptation decisions to maintain "absolute risk" at current levels by augmenting protective structures to maintain current expected annual damages. Abadie (2018) provide a similar assessment of future flood damage costs for 120 of the largest coastal cities by using expected shortfall as a risk measure, as opposed to (e.g.) expected annual damages. The use cases of no-adaptation and economically-efficient adaptation provide upper and lower bounds, respectively, on future total adaptation costs and damages, but do not represent the reality that many areas will likely pursue a strategy that falls somewhere in the middle. Thus, the results of these studies, including the present one, should be viewed as bounding scenarios, as opposed to specific predictions. Further, while many studies focus on the economic risks from coastal flooding, impacts can also be quantified in terms of the number of affected people living in flood-prone areas (e.g., Hinkel et al., 2014; Strauss et al., 2021). Indeed, as digital elevation models (DEMs) continue to evolve, estimates of numbers of coastal residents living on sea-level rise-implicated lands will evolve as well.

Finally, the original modeling experiments using CIAM (Diaz, 2016) characterized the impacts of uncertainty in future sea-level rise by examining the anticipated coastal adaptation costs and damages under the 5th, 50th, and 95th percentiles of global sea level projections. This method of using a few cases to characterize low-, moderate-, and high-risk scenarios has the advantage of being computationally inexpensive and conceptually simple. Previous studies have leveraged this computational efficiency in assessing flood risks (e.g., Hinkel et al., 2014; Lincke & Hinkel, 2018; Liuzzo & Freni, 2019; Tiggeloven et al., 2020). However, previous work has also found that using only a point estimate of an exogenous model input can miss substantial variation (Sieg et al., 2019) and multimodality (Brown et al., 2015) in the distributions of outputs of interest, which can also lead to underestimating the probability of damaging floods (Zarekarizi et al., 2020).

The risk of underestimating coastal adaptation costs and damages elevates the importance of interrogating how uncertainty propagates through CIAM to lead to changes in the distribution of future coastal adaptation costs, when uncertainty in sea-level rise is propagated through the model. For example, the 5th percentile among sea-level rise scenarios does not necessarily correspond to the 5th percentile among coastal adaptation cost scenarios. This is due to two main factors: i) sea-level rise does not occur uniformly around the world, and ii) CIAM contains nonlinearities and local factors related to adaptation costs and socioeconomic conditions. A more comprehensive, but often computationally infeasible, accounting of uncertainty will compute CIAM estimates for adaptation costs and decisions under full ensembles of sea-level rise uncertainty, as opposed to only a few benchmark percentiles.

Here, we address this issue by presenting an updated version of CIAM that accounts for limited decision-maker foresight in estimates of future sea levels. We use this new model version to present a set of experiments in which we characterize the impacts on coastal adaptation of propagating uncertainty in sea-level rise, uncertainty in socioeconomic conditions, and uncertainty in both. Finally, we use a global sensitivity analysis approach to determine the specific geophysical and socioeconomic uncertainties to which future coastal adaptation costs are most sensitive. In Section 2, we describe the modeling framework for sea-level rise and coastal adaptation decision-making and a set of experiments using this model. In Section 3.1, we present the results from a baseline set of experiments, decomposing adaptation costs across a set of socioeconomic and emissions pathways. In Section 3.2, we demonstrate the importance of a more complete accounting of uncertainty in future sea-level changes as opposed to taking only a few representative scenarios. In Section 3.3, we examine how uncertainties from geophysical and socioeconomic factors propagate to affect adaptation costs. In Sections 4 and 5, we summarize our findings and discuss the implications of these results.

## 2 Methods

### 2.1 Models

We use an open source modeling framework to relate radiative forcing scenarios to future coastal damages and adaptation decision-making. Our models are specifically chosen to provide a quantification of uncertainty for future coastal damages. We describe the basic model structure in greater detail in Sections 2.1.1 and 2.1.2, and a schematic is provided in Supporting Information (Figure S1).

#### 2.1.1 Coastal Adaptation

We use MimiCIAM v1.0.0 ("MimiCIAM"), an updated version of the original Coastal Impact and Adaptation Model (CIAM) by Diaz (2016), to compute adaptation costs at the level of individual coastal segments. These segments are from the DIVA database (Vafeidis et al., 2008). There are a total of 12,148 segments, partitioned so each segment has internally consistent physical characteristics. The median segment length is 18 km. MimiCIAM divides its time horizon into 40 or 50-year "adaptation periods" and runs with a 10-year time step. These adaptation periods are 2010-2050, 2050-2100, and 2100-2150. These periods were chosen for consistency with Diaz (2016) and to reflect the fact that 2050 is the time horizon at which sea-level projections begin to appreciably diverge, based on modeling assumptions and emissions scenario (Fox-Kemper et al., 2021). Each segment selects among three potential adaptation choices: protection (seawall construction), retreat (planned retreat from the coastline), or no adaptation (see Figure 1). For each segment, MimiCIAM computes the total costs of adaptation and damage for each combination of adaptation strategy and level (including no adaptation) as described in Diaz (2016). Total costs are aggregated over five categories: wetland loss, retreat/relocation costs, inundation (dryland loss), loss of property and life due to flooding, and construction costs (costs associated with building seawall protection). We provide further details about the MimiCIAM the model structure for computing these adaptation decisions and costs in Supporting Information (Text S2), but for a deeper discussion, see Diaz (2016).

Following the original CIAM implementation, the "optimal" strategy for each segment is chosen as the combination of strategy and level that minimizes the net present value (NPV) of the total expected cost. Distinct from the original CIAM implementation, we only consider the NPV over the first adaptation period (2010-2050) when choosing the optimal strategy, whereas the original model considers the NPV across the entire time horizon. Our approach more realistically mirrors real-world planning horizons for coastal adaptation. At the start of subsequent adaptation periods, the adaptation measures are updated to maintain the nominal level of defense over the course of that adaptation period. For example, suppose that building seawall defenses to guard against the 100-year extreme sea level is determined to be optimal for a given segment in the first adaptation period, in the year 2010. Then, when the next adaptation period begins in the year 2050, the height of these seawalls will be increased to account for the fact that sea-level rise will cause the 100-year return level to increase during the second adaptation period. This decision and adaptation process is summarized in Figure 2.

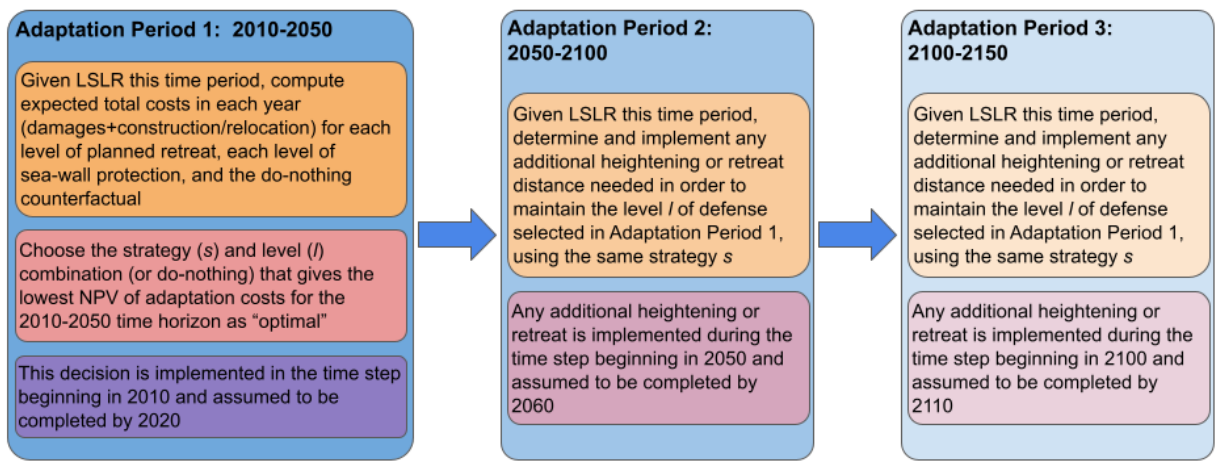

**Figure 2.** The MimiCIAM model decision-making process first chooses an adaptation strategy and level in the first adaptation period (2010-2050), which minimizes the net present value (NPV) of adaptation costs over that period, given the local sea-level rise (LSLR) projections. Then, in subsequent adaptation periods, the amount of protection or retreat is updated to maintain the chosen level of protection over the course of each adaptation period, assuming knowledge of future sea-level rise is limited to only the current adaptation period.

We run the MimiCIAM model starting in 2010 for two reasons. First, the CIAM initialization routine accounts for the dearth of information regarding initial adaptation states on a global scale by assuming an initial state of efficient present-day adaptation. This involves starting the model with no adaptation, and running a single time step (2010-2020). Second, we can use this common starting time to benchmark our results against those using "Original CIAM", the version as used in Diaz (2016). In the figures and data tables presented here, we subtract off the adaptation costs and damages from this reference period (2010-2020) because those costs are assumed to be in the past. In the experiments presented in this paper, we show results for hypothetical futures in which all segments follow a no-adaptation strategy and one in which all segments follow the least-cost adaptation strategy. These serve as practical bounds on the likely future total adaptation costs, but will of course miss the heterogeneity in adaptation measures pursued by coastal areas. As an example of the consequences of this exclusion, Hummel et al. (2021) find that protection in one coastal area can drive up damages from storms in other nearby areas.

Our new MimiCIAM model version is coded in the Julia programming language (Bezanson et al., 2017) for use within the Mimi integrated modeling framework (https://www.mimiframework.org/). The implementation of CIAM open source in the Mimi modeling framework is an important modeling advance to facilitate the use of MimiCIAM in climate change impacts studies, coupled to other climate and socioeconomic impacts models in a modular fashion (Moore et al., 2018; National Academies of Sciences, 2017).

### 2.1.2 Sea-level change

We examine the impacts of sea-level rise on estimates of future coastal damages using projections of global sea-level change from the Building Blocks for Relevant Ice and Climate Knowledge (BRICK) model (Wong, Rennels, et al., 2022a). This set of BRICK projections accounts for uncertainties in equilibrium climate sensitivity, ocean vertical heat diffusion, the modulation of effective radiative forcing by aerosols, and model parameters that govern the contributions to sea-level rise from glaciers and ice caps, thermal expansion, land water storage, the Greenland ice sheet, and the Antarctic ice sheet. Further details regarding BRICK may be found in Supporting Information (Text S6).

### 2.2 Updates and experiments

### 2.2.1 Single simulations

We conduct a series of initial simulation experiments to verify the match between Original CIAM and our new MimiCIAM. For these simulations only, we correct the model output in post-processing to have perfect foresight with respect to future sea levels over the full model simulation (see Supporting Information: Text S4 and Figure S3). This is made possible by including all of the potential adaptation options and their costs in the model output files. After confirming that MimiCIAM faithfully reproduces the Original CIAM results when using the same exogenous forcing, we make several model improvements. Below, we summarize the MimiCIAM changes relative to Original CIAM.

**Perfect versus limited foresight:** The Original CIAM model structure determines the least-cost adaptation decision by minimizing the NPV of total adaptation costs over the entire model simulation time horizon. This assumes that sea-level rise is known with perfect foresight. In MimiCIAM, we implement an update to the model so that the decision-maker has perfect foresight of future sea-level rise, but only over the course of a single adaptation period instead of the full model simulation. In MimiCIAM, we use the NPV of total adaptation costs over the first adaptation period (2010-2050) to determine the least-cost adaptation decision. From this point forward, MimiCIAM has limited foresight by default, unless otherwise explicitly stated.

**Updated gross domestic product and population data:** In MimiCIAM, we incorporate updated GDP and regional population scenarios via the Shared Socioeconomic Pathways (SSP) scenarios (Riahi et al., 2017). We show experiments using GDP and population data from the SSP pathways SSP1 and SSP5 as low- and high-end scenarios, and SSP2 and SSP4 as moderate pathways (O'Neill et al., 2016). Other population scenarios and projections are available and can be incorporated into MimiCIAM, but the SSPs are a commonly used tool for assessing climate change impacts and adaptation decision-making. Their implementation in MimiCIAM will facilitate comparisons with other similar work (e.g., Jevrejeva et al., 2018). In Supporting Information, we provide figures comparing the population and GDP scenarios, aggregated globally and for seven regions, as defined by the World Bank (https://datahelpdesk.worldbank.org/knowledgebase/articles/906519-world-bank-country-and-lending-groups): i) East Asia and Pacific, ii) Europe and Central Asia, iii) Latin America and the Caribbean, iv) Middle East and North Africa, v) North America, vi) South Asia, and vii) Sub-Saharan Africa (Figures S4 and S5). Note that the population and GDP scenarios shown in Figures S4 and S5 are only aggregated for the coastal segments and do not reflect totals for those regions or globally. Within DIVA, the population within a given segment is assumed to be distributed uniformly (Vafeidis et al., 2008). However, we stress that all segments are coastal and CIAM applies an adjustment to segments' per capita income to reflect the fact that more urban coastal areas are more likely to have higher income levels as compared to rural inland areas (Diaz, 2016) (see Supporting Information).

The present work employs the Global Land One-kilometer Baseline Elevation (GLOBE) DEM for parity with the original CIAM experiments, but we note that comparisons with other DEMs have varied results. Hinkel et al. (2014) find that the Shuttle Radar Topography Mission (SRTM) DEM yields estimates of exposure reduced by factors of 2-3, as compared to GLOBE DEM. On the other hand, Kulp & Strauss (2019) find that their more recent CoastalDEM yields estimates of exposure that are 2-3 times higher than the SRTM DEM. While intercomparisons of model results across DEMs are of course of interest, the experiments presented here center around model sensitivity. Specifically, we focus on examining how estimated adaptation costs are affected by endogenous uncertainties within the combined sea level-coastal adaptation model.

**Updated sea-level projections:** In MimiCIAM, we incorporate probabilistic projections of future sea-level rise that are updated to account for the deeply uncertain Antarctic ice sheet contribution to sea-level change (Wong, Rennels, et al., 2022b) and calibrated using a Bayesian model calibration approach (Vihola, 2012). We use the combined SSP-RCP (Representative Concentration Pathways) scenarios to represent low-end (SSP1-RCP2.6), moderate (SSP2-RCP4.5 and SSP4-RCP6.0), and high-end (SSP5-RCP8.5) scenarios (Moss et al., 2010; O'Neill et al., 2016, 2020; Riahi et al., 2017; Rogelj et al., 2018). In a supplemental experiment, we compare estimates of future coastal adaptation costs when MimiCIAM is run using the sea-level projections from Original CIAM (Kopp et al., 2014) to the cost estimates using these updated projections (Figure S6).

**New baseline model scenarios:** With these changes to assumed decision-maker foresight, socioeconomic forcing data, and sea-level projections, we conduct a set of model simulations using four SSP-RCP scenarios (SSP1-2.6, SSP2-4.5, SSP4-6.0, and SSP5-8.5). Following many previous studies in global coastal adaptation, for each scenario, we consider a baseline case of no adaptation, and a case where all coastal segments pursue the least-cost adaptation strategy. These results are explored in Section 3.1.

### 2.2.2 Ensemble simulations and propagation of uncertainty

In Section 3.2, we examine how a simplified quantification of uncertainty, which uses only a few percentiles of the distribution of future sea levels as opposed to a full ensemble, affects the resulting estimated distribution of coastal risks. Using the simplified approach, we take the BRICK simulations that give the 5th, 50th, and 95th percentiles of the distribution of GMSL in 2150 and use the corresponding sets of local mean sea-level rise as input for MimiCIAM.

For comparison, we also generate an ensemble of simulations from the coupled BRICK-CIAM model as follows. We sample 1,000 sets of BRICK model parameters from their joint posterior distribution (Wong, Rennels, et al., 2022b). We independently sample 1,000 sets of CIAM model parameters for six parameters: relocation cost as fraction of income, benchmark land value, population density elasticity of wetland value, income elasticity of wetland value, elasticity of the value of statistical life (VSL), and VSL multiplier on United States GDP. The distributions for these parameters are truncated normal distributions and are taken from the default distributions for generating Monte Carlo ensembles for the FUND integrated assessment model (Anthoff & Tol, 2014; Brander et al., 2006; Cline, 1992; Darwin et al., 1995; Diaz, 2016; Viscusi & Aldy, 2003). The specific distributions are given in Table S1. We perform a pair of sensitivity experiments using wider and narrower prior distributions for the CIAM socioeconomic parameters to examine the sensitivity of our results to changes in the widths of these distributions (see Figure S7). Supplemental testing also indicates that the results presented here are not sensitive to the choice of random seed for these Monte Carlo experiments.

We randomly pair each concomitant set of BRICK parameters with a corresponding set of CIAM parameters. We run MimiCIAM with all 1,000 of these sets of parameters to create an ensemble of estimates for future coastal damages. This constitutes a "soft coupling" configuration, in the parlance of Van Vuuren et al. (2012), or "loose, uni-directional coupling" to use the terminology of Srikrishnan et al. (2022). This loose, uni-directional form of coupling refers to the fact that information flows from BRICK to MimiCIAM, without feedbacks to BRICK (uni-directional), and that while we are sampling from the full posterior distribution of BRICK model parameters, the control volume for all practical purposes is restricted to the MimiCIAM model (loose coupling). These ensembles all use the updated population and GDP scenarios from the SSPs and limited decision-maker foresight. We include results using smaller ensemble sizes to verify that our results are not sensitive to changes in sample size (Figure S8).

We compute the NPV of total adaptation cost (including damages) for these ensembles across the 2010-2150 time horizon. We use a constant discount rate of 4% to match the Original CIAM model configuration. We also use a time horizon of 2150 for consistency, and because ending the model simulations at 2100 would lead to an overly optimistic characterization of uncertainty in adaptation costs towards the end of the century. This would occur because the hypothetical decision-maker in the year 2100, for example, would not need to plan for any additional sea-level rise beyond that year. We aggregate the total costs for each segment to obtain a single probability distribution for the expected total costs for the world as a whole. We also aggregate the costs into the seven World Bank regions (East Asia and Pacific; Europe and Central Asia; Latin America and the Caribbean; Middle East and North Africa; North America; South Asia; and Sub-Saharan Africa; see Supporting Information). For each region and for the whole world, we express the NPV of adaptation costs as a percent of the region's (world's) 2010 annual GDP. Monetary values are expressed in units of 2010 United States dollars.

### 2.3 Sensitivity Analysis

In Section 3.3, we compute "elementary effects" for each BRICK and MimiCIAM parameter using Method of Morris (Morris, 1991). Method of Morris is a one-at-a-time approach for sensitivity analysis and estimates the total global sensitivity of the model to each input parameter. We use this method for its computational efficiency relative to "all-at-a-time" methods (e.g., Sobol' analysis or Gini/permutation importances). However, this benefit of efficiency is balanced by a loss of interpretability because the estimated elementary effects do not represent, for example, a decomposition of variance. We use the means of the absolute values of the computed elementary effects ($\mu^*_i$ for parameter $i$, (Campolongo et al., 2007)) to characterize the sensitivity of total global adaptation costs to each model parameter. By examining the sorted magnitudes of all parameters' $\mu^*_i$ values we can characterize the BRICK and MimiCIAM parameters that drive uncertainty in coastal adaptation costs. Further, by also computing the standard deviation of the parameters' elementary effects ($\sigma_i$), we obtain an estimate of each parameter's level of interaction with other parameters (akin to higher-order sensitivity indices in Sobol' analysis) (Reed et al., 2022). We provide further details regarding the calculation of the means and standard deviations for the parameters' elementary effects in Supporting Information.

## 3 Results

### 3.1 Baseline scenarios and decomposition of adaptation costs

We examine the breakdown of globally-aggregated economic impacts of coastal flooding over the next century across the SSP-RCP scenarios (Figure 3). We provide tables with the model output used to generate these and other figures in the software and data repository accompanying this work (see Open Research statement).

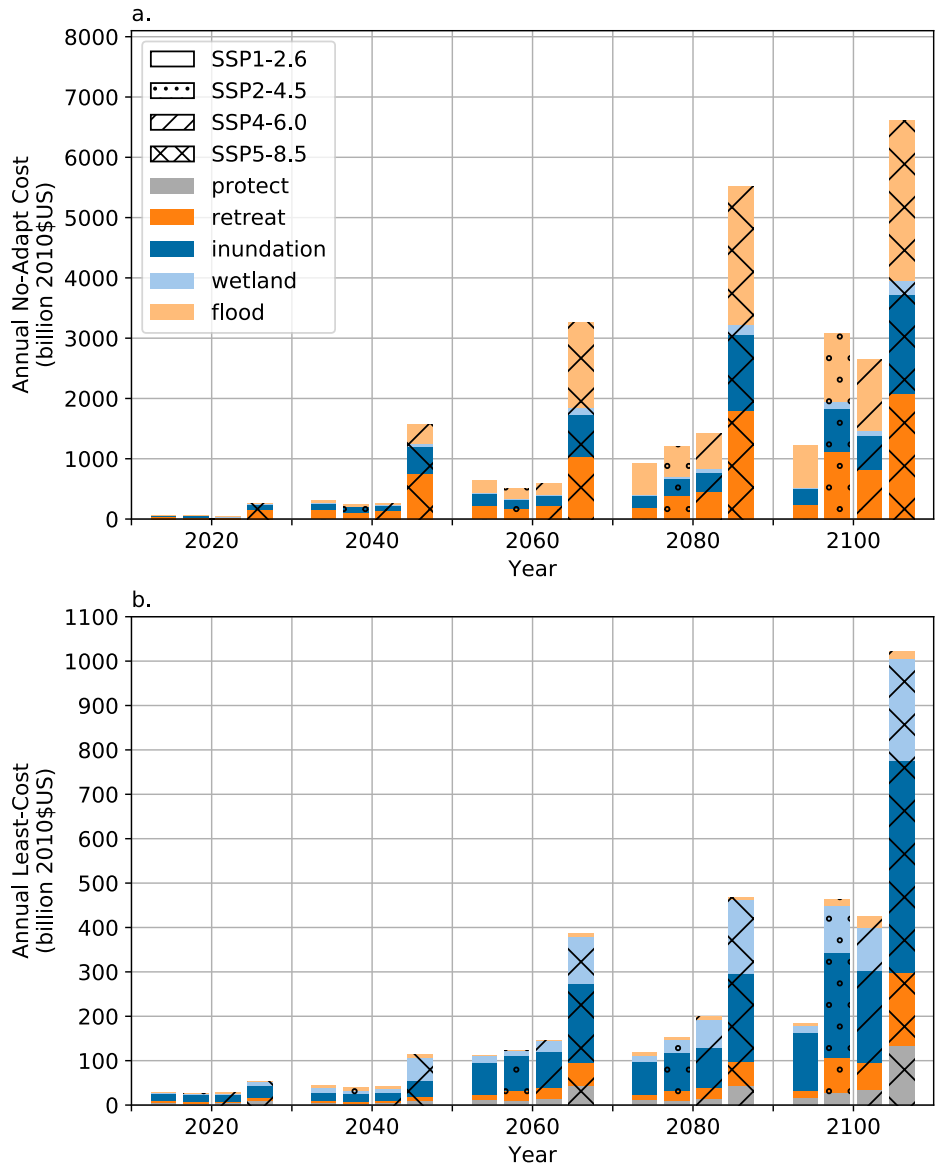

**Figure 3.** Comparison of total global adaptation costs and damages under SSP1-2.6 (no hatching, left-most boxes in each set of four), SSP2-4.6 (stippling, center-left), SSP4-6.0 (diagonal hatching, center-right), and SSP5-8.5 ("x" hatching, right), following (a) a no-adaptation scenario and (b) implementing least-cost adaptation by minimizing the NPV of total cost over each 40-50-year adaptation period.

These results demonstrate the trade-offs between mitigation/adaptation action and the types of costs incurred. For example, under the low-end SSP1-2.6 scenario, annual costs without adaptation reach about US$1.2 trillion by 2100 (Figure 3a). These costs are close to the US$1 trillion annual costs incurred under the high-end SSP5-8.5 pathway, but with least-cost adaptation implemented (Figure 3b). While the overall costs between these two scenarios are similar, they are distributed quite differently across damage categories. Of the US$1.2 trillion annual costs by 2100 under SSP1-2.6 without adaptation, US$699 billion are attributable to flood damages, including US$335 billion from population losses (Figure 3a). By contrast, with least-

cost adaptation in SSP5-8.5, even in this high-forcing scenario, losses of property and population from flooding total US$17 billion. Adaptation costs instead stem mostly from protection (US$135 billion), relocation (US$164 billion), and inundation (US$477 billion).

Coastal segments' reliance on protection to control overall costs is notably greater in SSP5-8.5 than the moderate SSP2-4.5 and SSP4-6.0 scenarios (Figure 3b). Of course, this is not surprising in absolute terms – US$135 billion in SSP5-8.5, compared to US$29 billion and US$36 billion in SSP2-4.5 and SSP4-6.0, respectively. However, as a percentage of overall costs, protection makes up about 13% of total adaptation costs by 2100 in SSP5-8.5, compared to 6 and 8% in SSP2-4.5 and SSP4-6.0, respectively. Retreat, by contrast, comprises between 14 and 17% of total adaptation costs across all three of those moderate or high-end scenarios, but only 8% of overall costs in SSP1-2.6. These results indicate the scenario-dependence of the efficacy of different strategies to manage coastal risk.

There are only slight differences between the two moderate scenarios, SSP2-4.5 and SSP4-6.0, attributable to the fact that the sea-level projections for RCP4.5 and RCP6.0 do not show substantial differences until after 2080 (Figure S2). Additionally, by 2040, the low- and moderate-forcing scenarios have not appreciably diverged, while SSP5-8.5 is notably higher. This is the case in both the counterfactual no-adaptation scenario (Figure 3a) and the least-cost adaptation scenario (Figure 3b). This is reflective of the sea-level projections (Figure S2); before 2060, there is little distinction between SSP1-2.6, SSP2-4.5, and SSP4-6.0.

### 3.2 Propagation of uncertainty

We compare estimates of the distribution of NPV of total least-cost adaptation costs over the 2010-2150 time horizon by using several alternative methods to propagate uncertainty. For each of the four SSP-RCP pathways considered, we first follow a common and simple approach of estimating the 5th, 50th, and 95th percentiles of the distribution of NPV of adaptation costs by using the corresponding percentiles from the distribution of sea-level rise scenarios (black dot and whiskers in Figure 4a-d). Against this three-percentile approach, we compare using a full ensemble of sea-level rise scenarios to compute a corresponding ensemble of estimated adaptation costs, then computing the appropriate percentiles from this full distribution (blue dot, whiskers, and distributions in Figure 4a-d). These first experiments consider uncertainty in future sea-level rise but keep the CIAM socioeconomic parameters fixed at their default values. In the moderate and high-end scenarios, the 95th percentile of the distribution of NPV is underestimated by the simulation using the 95th percentile from the ensemble of sea-level rise scenarios (Figure 4b-d). These underestimates of high-end risk range from about US$65 billion in SSP2-4.5 to US$231 billion in SSP4-6.0. The three-percentile approach also underestimates the width of the 5-95% range relative to the ensemble for NPV that considers a full distribution of sea-level rise, which can lead to overconfidence in adaptation decision planning. In order to help interpret some of these results, we also decompose these global costs across seven regions in the Supporting Information (Figure S9).

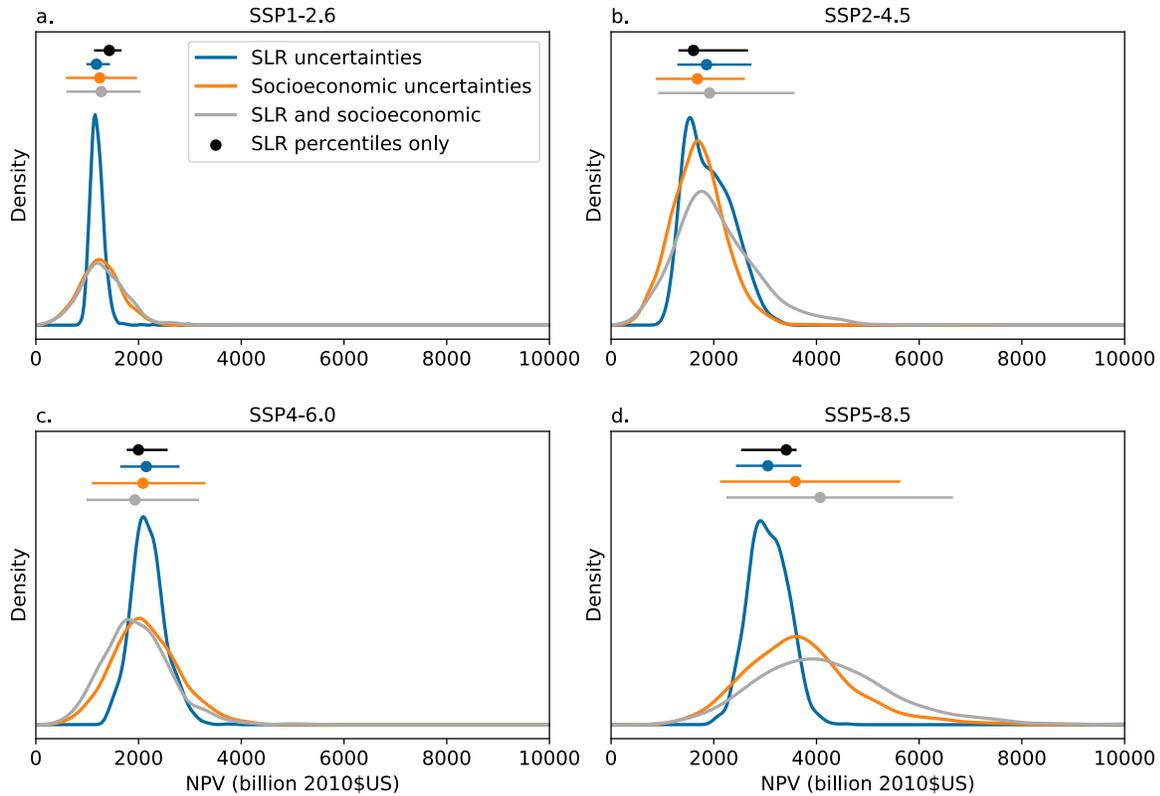

**Figure 4.** Kernel density estimates of least-cost NPV of total adaptation costs over 2010-2150 time horizon, aggregated for the whole world, under pathways SSP1-2.6 (a), SSP2-4.5 (b), SSP4-6.0 (c), and SSP5-8.5 (d). The dots and whiskers above the density estimates denote median and 5-95% ranges from the three-percentile approach (black), accounting for uncertainty in sea-level rise (blue), accounting for uncertainty in socioeconomic parameters (orange), and accounting for uncertainty in both sea-level rise and socioeconomic parameters (gray).

In these experiments, socioeconomic uncertainties dominate overall uncertainty in adaptation costs. When socioeconomic uncertainty is incorporated by varying the CIAM model parameters in addition to the BRICK ones, the 5-95% uncertainty range triples in width in SSP1-2.6 and SSP5-8.5 (Figure 4a, d, gray distributions compared to the blue distributions). In the moderate SSP2-4.5 and SSP4-6.0 scenarios, this uncertainty range nearly doubles (Figure 4b, c). Related to these substantial increases in the uncertainty ranges, incorporating both socioeconomic and sea-level uncertainties leads to sizable increases in the 95th percentile of damages in both moderate and high-end scenarios (Figure 4b, d). Specifically, in SSP2-4.5, when considering full distributions in either sea-level uncertainties or socioeconomic uncertainties, the 95th percentile of NPV of adaptation costs is US$2.6-2.7 trillion, and in SSP5-8.5, these high-end damages are US$3.7-5.6 trillion. However, when uncertainties are considered in both sea-level change and socioeconomic parameters, the high-end NPV of damages increases to US$3.6 trillion in SSP2-4.5 and US$6.6 trillion in SSP5-8.5. In all scenarios except SSP4-6.0, the median of the distribution of adaptation costs is also underestimated when both sea-level and socioeconomic uncertainties are not sampled. Thus, the simplification of holding uncertain parameters fixed at specific quantiles risks underestimating both expected as well as high-end tail risks. In the SSP5-8.5 pathway in particular, the median of the distribution of adaptation costs when both sea-level and socioeconomic uncertainties are sampled is substantially higher than the estimated median costs when only one or the other set of uncertainties is sampled (Figure 4d).

Importantly, when considering the prevalence of using several cases (often quantiles) to represent uncertainty (e.g., Hinkel et al., 2014; Lincke & Hinkel, 2018; Liuzzo & Freni, 2019; Tiggeloven et al., 2020), it is prudent to compare the distribution of damages from the full sampling of both sea-level and socioeconomic uncertainties (Figure 4, gray distributions) and the three-percentile approach (Figure 4, black dots and whiskers). In absolute terms, using the three-percentile approach underestimates the 95th percentile of the distribution of NPV of adaptation costs by US$371 billion (SSP1-2.6), $906 billion (SSP2-4.5), $611 billion (SSP4-6.0), and $3 trillion (SSP5-8.5). These differences constitute underestimates of 18, 26, 19, and 46% in the respective scenarios, relative to the more complete sampling of uncertainties.

### 3.3 Sensitivity Analysis

We employ the Method of Morris, a computationally efficient global sensitivity analysis approach, to assess which specific socioeconomic and sea-level uncertainties drive the sizable uncertainty in overall adaptation costs (Figure 5). Method of Morris estimates the relative importance and the level of interactions with other parameters of each uncertain model parameter by computing the mean of the absolute value and the standard deviation of the parameters' elementary effects ($\mu^*$ and $\sigma$, respectively). The specific values for the parameters' means and standard deviations of their elementary effects have no interpretation as (e.g.) a decomposition of variance. Consequently, only the ordering of the parameters matters in Figure 5, so no axis tick marks are given.

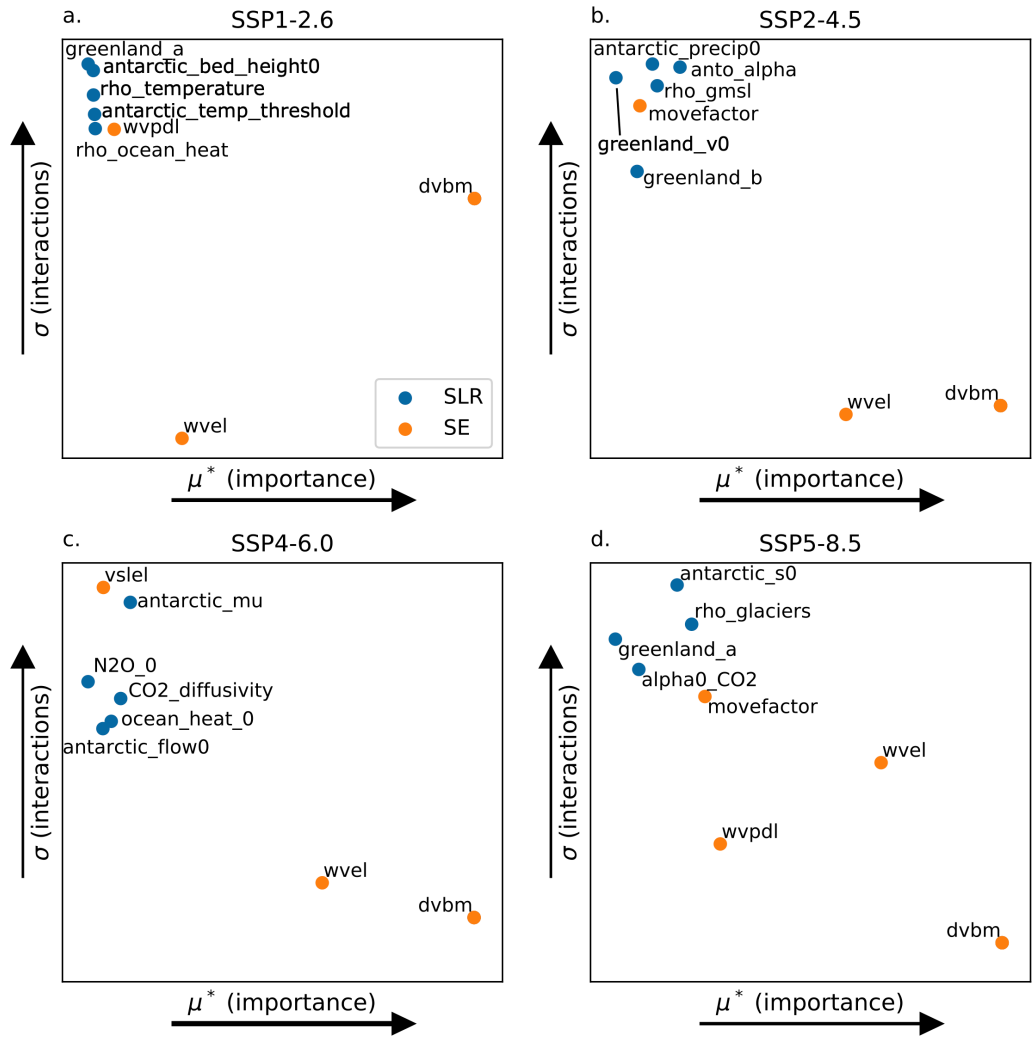

**Figure 5.** Estimated mean absolute value of the parameters' elementary effects (μ*) and the standard deviations of the parameters' elementary effects (σ) on the distribution of net present value of total adaptation costs and damages over the 2010-2150 period. Shown are scenarios (a) SSP1-2.6, (b) SSP2-4.5, (c) SSP4-6.0, and (d) SSP5-8.5. Socioeconomic (SE) parameters are shown with orange points and sea-level rise model (SLR) parameters are shown with blue points. For ease of viewing only the parameters with the eight largest distances from the origin are shown. No axis tick marks are provided because the specific values for μ* and σ are unimportant.

Across all SSP-RCP pathways, we find that the benchmark dryland value parameter (dvbm) and the income elasticity of wetland value (wvel) are consistently the parameters with the highest importance (Figure 5a-d; see Tables S1 and S2 for brief descriptions of the BRICK and CIAM parameters). There are 51 parameters related to the sea-level projections and six socioeconomic parameters within MimiCIAM being sampled. As expected from Figure 4, the socioeconomic parameters have a high presence among the parameters with the eight largest distances from the origin shown in Figure 5. The fact that the relocation costs as a fraction of income (movefactor) is an important parameter with relatively high interactions with other parameters in SSP2-4.5 and SSP5-8.5 is likely related to the prominence of relocation in those scenarios (17% and 16% of total costs by 2100, respectively; Figure 3). We also find that sea-level model parameters related to the Greenland and Antarctic ice sheets also contribute to uncertainty in overall damages through their strong interactions with other model uncertainties. In all scenarios, multiple ice sheet parameters are among those with the highest interactions ($\sigma$).

## 4 Discussion

In this work, we present an open-source and updated version of the Coastal Impact and Adaptation Model (MimiCIAM) and demonstrate its use in a set of experiments to propagate uncertainty and characterize sensitivity of coastal adaptation costs under a range of socioeconomic and climate forcing scenarios. We find that using only a set of percentiles of future SLR to characterize uncertainty in future coastal damages, as opposed to propagating uncertainty in the SLR model and socioeconomic parameters, leads to an underestimate in the 95th percentile NPV of future adaptation costs ranging from US$371 billion in SSP1-2.6 to US$3 trillion in SSP5-8.5 (Figure 4). Using only the 5th and 95th percentiles also underestimates uncertainty in future coastal adaptation costs by a factor ranging from just under 2 to more than 3.

The sizable increases in uncertainty in adaptation costs, particularly in the upper tails of the distributions for SSP2-4.5 and SSP5-8.5 (Figure 4) can be better understood by decomposing the global adaptation costs to the regional scale (Figure S9). Taking SSP5-8.5 as an example, the relative weight in the upper tails of these distributions is a reflection of the corresponding high-end adaptation costs seen in the Middle East and North Africa, Sub-Saharan Africa, and South Asia (Figure S9h, l, and x). All of these regions have relatively low population growth under SSP5 (Figure S4) but relatively high GDP growth (Figure S5). As many damages are related to GDP per capita (e.g., flood losses; Text S2), the assumed socioeconomic pathways drive higher risks for these regions, thereby raising the upper tail of global adaptation costs.

Further, this regional decomposition shows how the changes to the distribution of adaptation costs from incorporating socioeconomic and sea-level uncertainties vary by region. Sub-Saharan Africa, East Asian and the Pacific, and South Asia, for example, all see relatively little change in the mode of the distribution (under SSP5-8.5) when either sea-level or socioeconomic uncertainties are considered (Figure S9l, p, t, and x). However, when both sea-level and socioeconomic uncertainties are considered, the modes for these distributions of adaptation costs increase. By contrast, for the low forcing SSP1-2.6 pathway, the mode of the distribution of adaptation costs is relatively insensitive to incorporating the socioeconomic uncertainties (Figure S9, first column). The fact that the width of the distribution with both sea-level and socioeconomic uncertainties roughly matches the width with only socioeconomic uncertainties, however, indicates that for this low climate forcing pathway, the socioeconomic uncertainties dominate future adaptation costs.

The overall magnitude of the adaptation costs that we find in this study is in line with previous work. For example, Hinkel et al. (2014) find that seawall protection annual costs by 2100 total US$12-71 billion in 2005 US$. This amounts to roughly US$13-80 billion in 2010 US$, as compared to our range across the SSP-RCP scenarios of US$16-135 billion annual costs in 2100 (converted from 2005 to 2010 (June) dollars using U.S. Bureau of Labor Statistics Consumer Price Index Inflation Calculator: https://data.bls.gov/cgi-bin/cpicalc.pl (accessed 18

June 2022)). Our finding of higher damages is attributable to our use of updated sea-level change scenarios. This explanation is supported by the similarities between the studies' low-end estimates of annual costs (US$13 versus 16 billion), but more sizable differences in high-end damages (US$80 versus 135 billion), driven by higher high-end sea-level rise under RCP8.5 in our updated sea-level projections (see Figure S2).

In the no-adaptation counterfactual scenario, we estimate about US$2.4 trillion in annual expected damages under SSP5-8.5 in 2050. If both sea-level and socioeconomic uncertainties are not considered, these median damages are underestimated by up to US$1 trillion (Figure 4d). This median of US$2.4 trillion is slightly higher but still in agreement with the result of Abadie et al. (2020) giving a median anticipated damages of US$1.6 trillion under RCP8.5, with high-end damages (mean loss in the upper 5% tail) of US$3.1 trillion. That our result is notably higher than their median is not surprising, given that Abadie et al. (2020) considered a set of 136 major coastal cities, while our study considers all of the world's coastlines. However, we anticipate that the majority of losses are comparable, as most economic damage will be centered around the major metropolitan coastal areas that are included in both studies. Further, the present work extends previous studies by attributing uncertainty in these future adaptation costs and damages to socioeconomic uncertainties, in particular, the value of land being abandoned to encroaching sea levels.

These results are, of course, subject to a number of important caveats. We find that the socioeconomic uncertainties dominate the uncertainty in adaptation costs (Figures 4 and 5). However, it is important to keep in mind that the distributions of BRICK model parameters have been calibrated using Bayesian methods, whereas the socioeconomic parameters are being sampled from prior distributions (Table S1). Due to a lack of data with which to calibrate global estimates for these parameters, it is not feasible at this time to implement a similar Bayesian calibration scheme for the socioeconomic parameters in MimiCIAM. Thus, it is not surprising that these parameters contribute sizable uncertainties to the distribution of adaptation costs and, relatedly, the distributions of adaptation costs are somewhat sensitive to the chosen widths of these prior distributions (Figure S7). However, changing the scale parameters for these distributions by a factor of 2 in our experiments constitutes substantial changes to the shapes of these prior distributions. Further, our estimates for the distribution of adaptation costs as a percentage of global GDP are in rough agreement with the results of Jevrejeva et al. (2018), which indicates that the chosen distributions are, at least to first order, plausible.

There are also several key limitations of the modeling framework, including model structural uncertainties and assumptions. First, the estimated adaptation costs and damages presented here are conditioned on a single set of sea-level projections from just one of many possible sea-level models. As such, uncertainty in future adaptation costs due to sea-level model structural uncertainty is not represented in our results. We anticipate that model structural uncertainty, for both sea-level and coastal adaptation models, plays an important role in uncertainties in future coastal adaptation costs. For example, Ruckert et al. (2019) find substantial structural uncertainty among projections of local sea levels for Norfolk, Virginia, USA, including both the magnitude and timing of increases in local sea level. This uncertainty in coastal hazard likely translates to additional uncertainty in coastal risk and adaptation costs, although further work is needed. Second, the MimiCIAM model structure does not explicitly account for *in situ* adaptation strategies such as beach nourishment or the elevation of existing structures (e.g., by building a new level at the top of a structure and abandoning the bottom level). These adaptation measures may be roughly categorized as equivalent to some level of protection via managed retreat or seawall construction, but this still misses the potential for heterogeneity within a coastal segment. Relatedly, MimiCIAM also does not incorporate potential feedbacks between adaptation strategies and their efficacies in neighboring segments (Hummel et al., 2021), nor does the model's accounting of expected costs include non-economic costs such as cultural heritage and sense of place.

Additionally, the decision-making process in MimiCIAM assumes that the decision-maker has perfect foresight of future sea-level rise over the course of each adaptation period (40-50 years), and that they will implement the adaptation strategy and level that minimizes the NPV of total adaptation costs over the course of each adaptation period. In practice, this least-cost case and the counterfactual no-adaptation scenario may serve as bounds on the efficacy of strategies to adapt to coastal hazards. This, of course, does not account for the potential for partial and heterogeneous adaptation within a single coastal segment, delayed adaptation efforts, and imperfect maintenance of protective structures. All of these factors, as well as scenarios intermediate to the SSP-RCP pathways examined here, will raise potential adaptation costs above the least-cost scenarios from MimiCIAM. The MimiCIAM model structure also does not allow decision-makers to alter the chosen adaptation strategy after the first adaptation period. Thus, there is a need for more sophisticated decision support frameworks to inform dynamic on-the-ground adaptation for specific locations (e.g., Haasnoot et al., 2020; Haasnoot et al., 2013). These model structural uncertainties also likely contribute appreciably to uncertainty in future coastal adaptation costs. On a similar note, MimiCIAM does not account for potential governmental policies regarding adaptation efforts, including but not limited to disaster relief programs, flood insurance programs, and home buyouts for proactive relocation. Indeed, the results presented here (Figure 3), as well as the study of Lincke & Hinkel (2021), demonstrate the importance of proactive relocation to mitigate loss of life and property. Finally, the estimates of storm surge return levels used in MimiCIAM do not account for potential changes in the probability distributions or frequencies of extreme sea levels.

## 5 Conclusions

Aforementioned caveats notwithstanding, the set of experiments presented here demonstrates the dangers of underestimating high-end coastal risk when only using a loose coupling between the model for future sea levels and the model for coastal impacts and adaptation. Further, by employing the Mimi coupled modeling framework, the MimiCIAM model can serve as a starting point to incorporate coastal adaptation and risks into broader-scale efforts for modeling climate and socioeconomic impacts. Our simulations indicate that characterizing uncertainty in sea level by using only low, moderate, and high cases can dramatically underestimate future coastal damages relative to using an ensemble approach to propagate sea-level uncertainty (Sec. 3.2). We also find that the percentiles-only approach underestimates uncertainty in future coastal adaptation costs by a factor ranging from nearly 2 to more than 3. This misrepresentation of the upper tail and uncertainty in future sea-level rise and coastal impacts can have measurable consequences due to overconfidence in adaptation planning. Decision-maker overconfidence can lead to a failure to hedge against the most severe outcomes. This, in turn, can manifest as failed seawall protection or a population being forced to retreat reactively as their homeland becomes inundated over time.

This work considers both a future without adaptation, and one in which decision-makers execute a cost-minimizing adaptation strategy in 40-50-year periods of foresight. Our results highlight that decision-makers can – and should – use investments in near-term adaptation efforts to mitigate higher long-term costs from loss of life and property. Through our propagation of uncertainty experiments, we have demonstrated the importance of accounting for the contributions from both geophysical and socioeconomic uncertainties to the uncertainty in future coastal risk. Neglecting either can have severe consequences in terms of overconfidence and underestimating high-end risks, resulting in inefficient or ineffective adaptation measures to protect coastal populations and property.

## Acknowledgments

Contributions from T.E.W. were supported in part by the National Science Foundation under Award No. DMS 2213432. Contributions from C.L. were supported by the National Science Foundation Graduate Research Fellowship Program under Grant No. DGE 1752814. Contributions from H.S. were supported by an Inclusive Excellence grant from the Howard Hughes Medical Institute. Any opinions, findings, and conclusions or recommendations expressed in this material are those of the authors and do not necessarily reflect the views of the funding agencies. We thank Corinne Hartin, Klaus Keller, Radley Powers, Vivek Srikrishnan, and Ryan Sriver for valuable inputs.

## Open Research

The MimiCIAM v1.0.0 model code is available as a Julia package and may be found on GitHub at https://github.com/raddleverse/MimiCIAM.jl. The code used in this work to run the model and analyze the output is available on GitHub at https://github.com/raddleverse/CIAM_uncertainty_propagation. These codes are also available with the output files on Zenodo at https://zenodo.org/record/6835363.

T. E. Wong[1], C. Ledna[2], L. Rennels[2], H. Sheets[1], F. C. Errickson[3], D. Diaz[4], and D. Anthoff[2]

[1] School of Mathematical Sciences, Rochester Institute of Technology, Rochester, New York USA.

[2] Energy and Resources Group, University of California Berkeley, Berkeley, CA, USA.

[3] School of Public and International Affairs, Princeton University, Princeton, NJ, USA.

[4] Energy Systems and Climate Analysis Group, Electric Power Research Institute, Palo Alto, CA, USA.

**Contents of this file**

Text S1 to S6
Figures S1 to S11
Table S1 to S2

## Introduction

This Supporting Information includes some additional guidance for those who would like to use the MimiCIAM implementation of the Coastal Impact and Adaptation Model (CIAM), initial validation against the previous version of CIAM, descriptions of the main damage functions within CIAM, a supplemental experiment to examine sensitivity to the chosen storm surge exposure data set, and some tables and figures to support and expand on the results presented in the main text. The additional user guidance is meant to help distinguish between the various software components of our work. Specific details for replication of these results and running new cases with the MimiCIAM model are given in the code repositories online and linked in Text S1 below.

The Dataset files that we provide include all of the raw or processed data (in CSV form) that is necessary in order to reproduce the figures shown in the main text and the Supporting Information figures. The Monte Carlo sampling procedure is necessarily random, so if the code is re-run from scratch then the exact quantiles may differ slightly from those presented in the main text. However, Figure S8 suggests that they will not differ substantially.

### Text S1.  Guidance for users of MimiCIAM

Three repositories hold the bulk of support for this paper, starting with Mimi.jl, which provides the framework and baseline support for the implementation of the CIAM model (Diaz, 2016) in the second repository, MimiCIAM.jl. Both Mimi and MimiCIAM are available as packages that are imported in the third repository, which is specific to this project. The third repository detailed below leverages MimiCIAM.jl to perform the work and experiments for this paper, and provides support for replication and exploration of our results. The following is a brief overview of the purpose and contents of each of these repositories. More detailed user information may be found within the individual repositories linked below.

**Mimi.jl** (https://www.mimiframework.org): The Mimi project is a Julia package that provides a component model for integrated assessment models. Much of the work aligns with the aforementioned National Academies of Sciences (National Academies of Sciences & Medicine, 2017) recommendation for a flexible, modular, uniform computational platform for integrated assessment models.

**MimiCIAM.jl** (https://github.com/raddleverse/MimiCIAM.jl): MimiCIAM is a Julia package leveraging the Mimi.jl framework to produce a Julia implementation of the CIAM model adapted from Diaz (2016). The README for this repository contains details on running the default version of the model, looking at the results, and some of the options for modifying the parameterization and underlying data such as the local mean sea level rise projections as discussed next. Details on other inputs are available upon request if the information in the README does not answer further questions.

CIAM input data for projections of local mean sea levels are provided as a comma-separated value (CSV) file. Each column corresponds to a CIAM coastal segment and has a column header (first row) containing the name of that segment. Each row corresponds to the local annual mean sea-level change relative to the start of the model simulation period, for each model time step. An input argument gives the CIAM user the option to provide a list of segments that is a subset of the full set of 12,148 coastal segments, and run CIAM with only this subset. The model codes that accompany this work include a file that provides the latitude and longitude coordinates for each coastal segment, a script that runs CIAM for a given subset of segments (if provided), and writes CSV model output files to save the costs and optimal adaptation decisions for each segment for each time step.

**CIAM_uncertainty_propagation** (https://github.com/raddleverse/CIAM_uncertainty_propagation): This repository holds the scripts for this paper, which rely heavily on the infrastructure in the two preceding repositories, MimiCIAM.jl and Mimi.jl. These scripts allow for replication of paper results, including two primary types of experiments: (1) the baseline MimiCIAM simulations and (2) the Monte Carlo Ensembles. The README for this repository contains details on running these on a local machine, and/or running analysis through provided notebooks. Details are available upon request if the information in the README does not answer further questions.

**Text S2. CIAM adaptation strategies and cost estimates**

In the event that a coastal segment in CIAM becomes inundated due to local mean sea-level rise, that segment will retreat reactively (in contrast to planned, or proactive, retreat). For the retreat and protection strategies, segments can choose a defense level from among the 10, 100, 1,000, and 10,000-year (presumed maximum) storm surge return periods. Retreat for the 1-year return period is also an option. These protection levels are calculated for each segment based on generalized extreme value distributions, and reported in the DIVA database (Vafeidis et al., 2008). In light of the biases inherent in any global-scale database of storm surge return levels (Hunter et al., 2017; Muis et al., 2017), we conduct a supplemental experiment (see Supporting Information Text S3) to assess the degree to which our estimated coastal adaptation costs and damages are influenced by use of the DINAS-COAST surge exposure data set within DIVA, following the original CIAM implementation. Also following the original formulation of Diaz (2016), reactive retreat is assumed to cost five times the same level of planned retreat.

In CIAM, costs are decomposed into five categories: wetland loss, retreat/relocation costs, inundation (dryland loss), loss of property and life due to flooding, and construction costs (costs associated with building and maintaining seawall protection). Here, we briefly review the structural assumptions for estimating these costs for an arbitrary time step at an arbitrary coastal segment. For a deeper discussion, the reader is directed elsewhere in the literature, as these have been discussed at length in other works (e.g., Diaz (2016), Vafeidis et al. (2008), Hinkel and Klein (2009), Hinkel et al. (2014)). This description closely mirrors the Supplementary Material (Sec. 2) of Diaz (2016), which will be of high relevance to interested readers.

**Protection costs** are computed from a linear contribution from coastline length, a quadratic contribution from seawall height, and a location-dependent annual cost to occupy the land and maintain the seawalls. This equation is:

$$ProtectionCost = l \cdot pc \cdot (H^2 + mcH) + l \cdot lv \cdot 1.7H,$$

where *l* is the length of coastline, *pc* is a country-specific protection construction cost, *mc* is the maintenance cost, *H* is the height of the seawall protection, and *lv* is the occupied land value. The factor of 1.7*H* represents the width of the seawall, stemming from an assumption within CIAM that seawalls have a 60-degree slope on each side. The "seawall" protection is meant to generalize the specific types of protective measures that are appropriate to individual geographies of the coastal segments. Reference costs for this generic protection is computed from a review of average dike and seawall costs (Hillen et al. 2010) and a database of national construction and labor cost indices (World bank International Comparison Program, 2011).

**Retreat costs** include the cost of relocating mobile capital and population. The losses from inundation of the evacuated area are included in the next section. Retreat costs are given by:

$$RetreatCost = \theta_L \sigma_L area(R - R_o) + (\theta_K + dc)\sigma_K area(R - R_o),$$

where $\theta_L$ and $\theta_K$ are cost coefficients for retreat per unit population and mobile capital (respectively), $\sigma_L$ and $\sigma_K$ are densities of population and mobile capital, *dc* is a demolition cost for immobile capital. Within CIAM, it is assumed that one-fourth of capital is mobile and that demolition costs for immobile capital are 5% of the capital's value. As retreat can be proactive or reactive, following Diaz (2016), we assume that reactive retreat costs five times as much as proactive retreat. Segments are also assumed to be capable of relocating capital and population within the segments' borders, and with full productivity upon relocation.

**Inundation costs** represent the losses of unprotected land that falls below local mean sea level. Inundation costs are computed as:

$$InundationCost = lv\, area + (1 - \delta)\sigma_K area,$$

where *area* is the area of newly-inundated land and $\delta$ is a depreciation rate parameter. It is assumed that mobile capital stocks (25% of all capital stocks) are reactively relocated when inundated, but immobile capital stocks (75%) are abandoned. However, if a segment pursues proactive retreat as its adaptation strategy, this foresight can avert significant losses. This is represented by the $\delta$ depreciation parameter, which is between 0 and 1 (Yohe et al., 1995). Interior land value comes from the Global Trade Analysis Project (Baldos and Hertel, 2012). It is assumed that land values increase with increases in income and population density (Yohe et al., 1999).

**Wetland costs** account for the impacts of sea-level rise and coastal adaptation decisions on wetlands. Wetland costs are assumed to be quadratically related to the rate of sea-level rise if segments either do not adapt or follow a proactive retreat strategy. If segments construct protective seawalls, then wetland services are assumed to be totally lost because wetlands no longer have inland space to migrate away from advancing sea levels. Wetland costs are computed as:

$$WetlandCost = \begin{cases} wv\ area \left(\frac{\frac{dSLR}{dt}}{\lambda}\right)^2 & \text{if Retreat or No Adaptation and } \frac{dSLR}{dt} < \lambda \\ wv\ area & \text{if Protect or } \frac{dSLR}{dt} \geq \lambda \end{cases},$$

where *wv* represents the annual value of wetland services (Brander et al., 2006), *dSLR/dt* is the rate of sea-level rise, λ is a threshold for wetland migration, and *area* is the inundated wetland area (Hoozemans, 1993; Spalding, 1997). Wetland value is also assumed to increase with income and population density.

**Flood costs** are computed as the expected damages over the distribution of extreme sea levels, *s*:

$$FloodCost = E[Damage(s)] = \int_A^{s_{max}} Damage(s) f(s) ds$$

where *f(s)* is the probability density function for extreme sea level *s*, *A* is the current adaptation level for this segment, $s_{max}$ is a maximum plausible surge level for this segment, and *Damage(s)* gives the flood damages associated with the incidence of extreme sea level *s*. *f(s)* is assumed to follow a generalized extreme value distribution (Diaz, 2016). We conduct sensitivity experiments to compare updated surge exposure data sets (Muis et al., 2016) against this original surge exposure data (see Text S3). *Damage(s)* is given by:

$$Damage(s) = (1 - \rho) \int_{x_{min}}^{lslr+s} area'(x) \cdot \left(\sigma_K\ \phi(h(x)) + \sigma_L\ \mu\ VSL\right) dx,$$

where $x_{min}$ is the lower bound of elevation for this segment, ρ is a national resilience index parameter (which is tied to national gross domestic product), *x* is vertical elevation, *lslr* is the change in local mean sea level from the start of the model simulation, *h* is flood water height, ϕ(*h*) is a depth-damage function (Hinkel et al., 2014), µ is a flood mortality factor (Jonkman and Vrijling, 2008), and *VSL* is the national value of a statistical life for the country in which the given segment is located. *VSL* is based on national per capita income (216 times per capita income), which is consistent with how (for example) the FUND integrated assessment model computes the value of a statistical life (200 times per capita income) (Cline, 1992).

We note that there are, of course, many parametric uncertainties (e.g., VSL or the flood mortality factor) and structural uncertainties (specific data sets used, the form of extreme value distribution for surge exposure, or the parameterization of each component of overall damages). This work is not meant to be an attempt to minimize or constrain these uncertainties. Rather, it is our aim to characterize how these uncertainties propagate through the coupled geophysical (sea-level change) and human (coastal risk) system. This characterization of uncertainty is conditioned on a particular set of structural and parametric assumptions that are consistent with numerous previous studies spanning the previous decade of research. These uncertainties, while potentially reducible, are inherent and unavoidable. No study can reduce them to zero, which elevates the importance of improving our understanding of how they influence our characterizations of coastal risk.

**Text S3. Sensitivity to surge exposure data set**

We conduct a supplemental experiment to assess the sensitivity of the modeled adaptation costs to the specific data set used for storm surge return levels for each CIAM coastal segment. The Original CIAM model uses the DINAS-COAST ("DC") surge exposure data as provided by the DIVA database (Vafeidis et al., 2008). We compare the DC data set to two alternative surge data sets, based on the Global Tide and Surge Reanalysis (GTSR) data set (Muis et al., 2016). We construct one data set that applies a multiplicative bias correction to the DC surge level data. This bias factor is equal to the ratio of the 100-year return level from GTSR to the 100-year return level from DC for each particular coastal segment. Then, we multiply each return level from the DC data set by this factor. Thus, in this "GTSR/DC" data set, the 100-year return level matches those from GTSR exactly, and the other return levels (10-year and 1000-year) are stretched/contracted to account for biases (e.g., that DC return level distributions are too high and too flat (Hunter et al., 2017)). We also compare a GTSR-based data set for each segment by using the nearest GTSR data point to each DIVA segment's centroid. The GTSR data (GTSR, 2016) does not include the "1-year" return level (annual maximum high tide level (Diaz 2016), so we use the GTSR/DC return levels for that datum.

To verify that these data sets produce surge exposure levels that reflect the biases relative to one another that have been previously noted in the literature (e.g., Muis et al., 2017), Figure S10 shows the relationship between the GTSR and the DC 100-year return levels. Muis et al. (2017) found a global mean bias for DC relative to tide gauge data of 0.55 m, and a bias for GTSR of -0.19 m. From the linearity of expectation, the mean bias of DC relative to GTSR should then be 0.74 m. DC surge levels are higher than our GTSR data set, with a mean difference, *DC-GTSR*, of 0.748 m, in tight agreement with the findings of Muis et al. (2017).

The distributions of the total global net present value of total adaptation costs from 2010-2150 (with a constant 4% discount rate) are not sensitive to the choice of specific storm surge exposure data set (Figure S11). This is potentially counterintuitive based on the apparent differences in the storm surge data sets (Figure S10), but is explained by the CIAM model structure (both in Original CIAM, and retained in our MimiCIAM implementation). Specifically, the first model time step (2010-2020) is a "calibration" or "spin-up" period. This accounts for the relative sparsity of detailed information about coastal adaptation infrastructure on a global scale. In this calibration period, it is assumed that each segment adapts efficiently (least-cost) to the storm surge hazard over this period. The storm surge distributions for each coastal segment are assumed to be stationary (i.e., return levels are the same throughout the model simulation time period). Thus, the potentially increased or decreased adaptation costs are felt in the first (calibration) time step, and subsequent increases in flood risk are driven entirely by sea-level rise. As noted in the main text, we have subtracted the costs/damages from this calibration period from the results in subsequent time steps.

**Text S4. Model validation and limited decision-maker foresight**

In a first experiment, we verify that the new MimiCIAM model, coded in Julia, matches the model output from Original CIAM, which was coded in GAMS (Figure S3, no hatch). We post-process the model output to be consistent with the perfect foresight of future sea-level rise that is assumed in Original CIAM (Figure S3, "x" hatch). We also compare the new version of MimiCIAM with limited decision-maker foresight (Figure S3, stippling). As expected, in the no-adaptation scenario, all three versions match (Figure S3a). With least-cost adaptation, for the first adaptation (2010-2050) annual total adaptation costs are slightly lower with limited foresight (Figure S3b). This is because perfect foresight of end-of-century sea-level rise leads to higher levels of protection in the Original CIAM and perfect foresight MimiCIAM cases. This is evident in the relatively lower protect costs in the limited foresight MimiCIAM case, and slightly higher inundation damages (Figure S3b). In subsequent adaptation periods (2050-2090, and 2100 and beyond), higher optimal adaptation costs in the limited foresight case are driven by inundation (dryland loss) and proactive retreat.

In the experiments presented in the main text, we employ GDP and population pathways that follow the SSPs (Riahi et al., 2017) and sea-level projections that use an updated semi-empirical sea-level model (Wong et al., 2022a). Relative to the limited foresight MimiCIAM baseline (Figure S3, stippling), we show under SSP5-8.5 the incremental changes from each of those two updates (Figure S6).

**Text S5. Calculation of elementary effects for sensitivity analysis**

We use Method of Morris (Morris, 1991) to compute elementary effects ($EE_{i,j}$) for each BRICK and MimiCIAM input parameter, indexed by $i$ = 1, 2, ..., 57, and for a number of trajectories $r$, indexed by $j$ = 1, 2, ..., $r$. If $M(\mathbf{x})$ denotes the model output of interest (here, net present value of total adaptation costs) when using input parameter vector $\mathbf{x}$, the model sensitivity to parameter $i$ is estimated as the mean of the absolute value of the $EE_{i,j}$ as:

$$S_i = \mu_i^* = \frac{1}{r}\sum_{j=1}^{r} |EE_{i,j}| = \frac{1}{r}\sum_{j=1}^{r} \left| \frac{M([x_1, x_2, ..., x_{i-1}, x_i + \Delta_{i,j}, x_{i+1}, ..., x_{57}]) - M([x_1, x_2, ..., x_{57}])}{\Delta_{i,j}} \right|, \quad (1)$$

where $r$ is the number of trajectories used in the estimate.

We estimate the standard deviations of the parameters' elementary effects as:

$$\sigma_i = \sqrt{\frac{1}{r}\sum_{j=1}^{r}(EE_{i,j} - \overline{EE_i})^2}, \quad (2)$$

where $\overline{EE_i}$ is the mean of the elementary effects for parameter $i$.

**Text S6.  Further details regarding the BRICK sea-level rise model**

Our version of BRICK is mechanistically identical to the version used by Vega-Westhoff et al. (2019), but implemented in the Julia programming language and written to comply with the Mimi modeling framework. The model takes as input a radiative forcing scenario and a set of model parameters. As output, BRICK yields changes in future global mean sea level (GMSL), and its contributions from glaciers and ice caps, thermal expansion, land water storage, the Greenland ice sheet, and the Antarctic ice sheet. The Antarctic ice sheet model subcomponent of BRICK includes a parameterization to represent the fast dynamical ice loss associated with marine ice sheet and ice cliff instabilities (DeConto et al., 2021; Kopp et al., 2017; Le Bars et al., 2017; Wong, Bakker, & Keller, 2017). Sea-level fingerprints from Slangen et al. (2014) are used to downscale the contributions to global mean sea-level changes to their effects on local mean sea-level change. This local mean sea-level change serves as input to CIAM.

BRICK is a semi-empirical model designed to be flexible and efficient, and thus usable in formal model calibration frameworks. The sea-level projections from (Wong, Rennels, et al., 2022b) were calibrated using a Bayesian model calibration approach that accounts for uncertainty in the model parameters, heteroskedastic uncertainty in the observational data, and autocorrelation in the time series of model-data residuals. The BRICK projections of GMSL are consistent with the estimates of future GMSL change from the Intergovernmental Panel on Climate Change Sixth Assessment Report (IPCC AR6; (Fox-Kemper et al., 2021; their Table 9.8)). Specifically, there is good agreement between the AR6 Marine Ice Cliff Instability (MICI)-based estimates of GMSL change and the BRICK projections (see Figure S2), which is expected given the treatment of the Antarctic ice sheet in BRICK noted above.

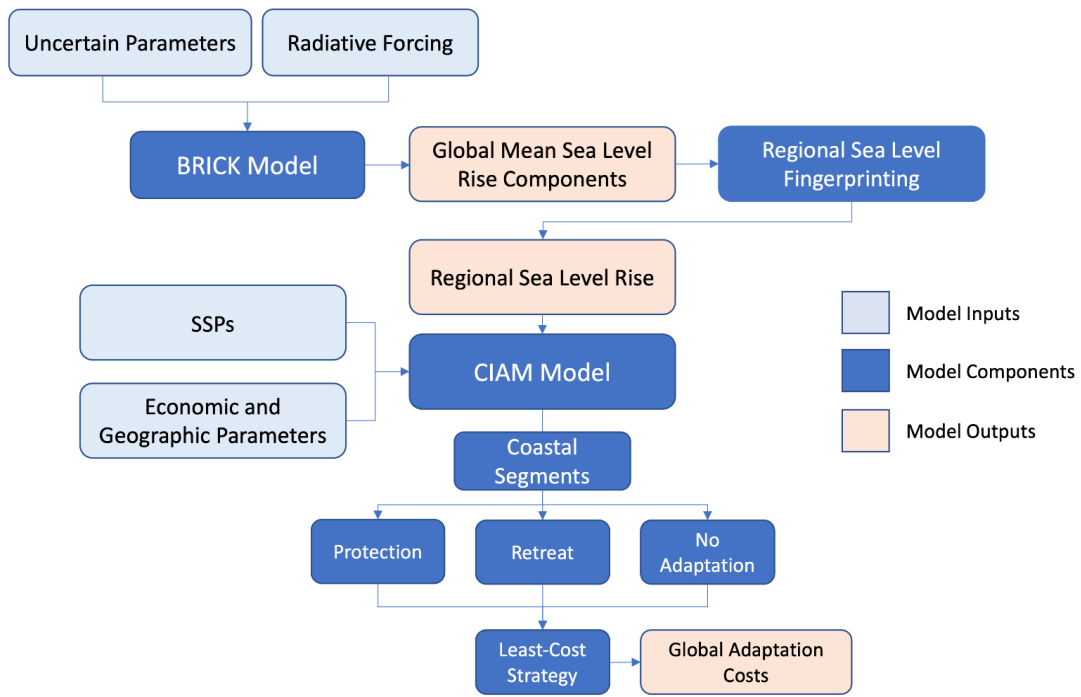

**Figure S1.** Coupled model structure for BRICK-CIAM.

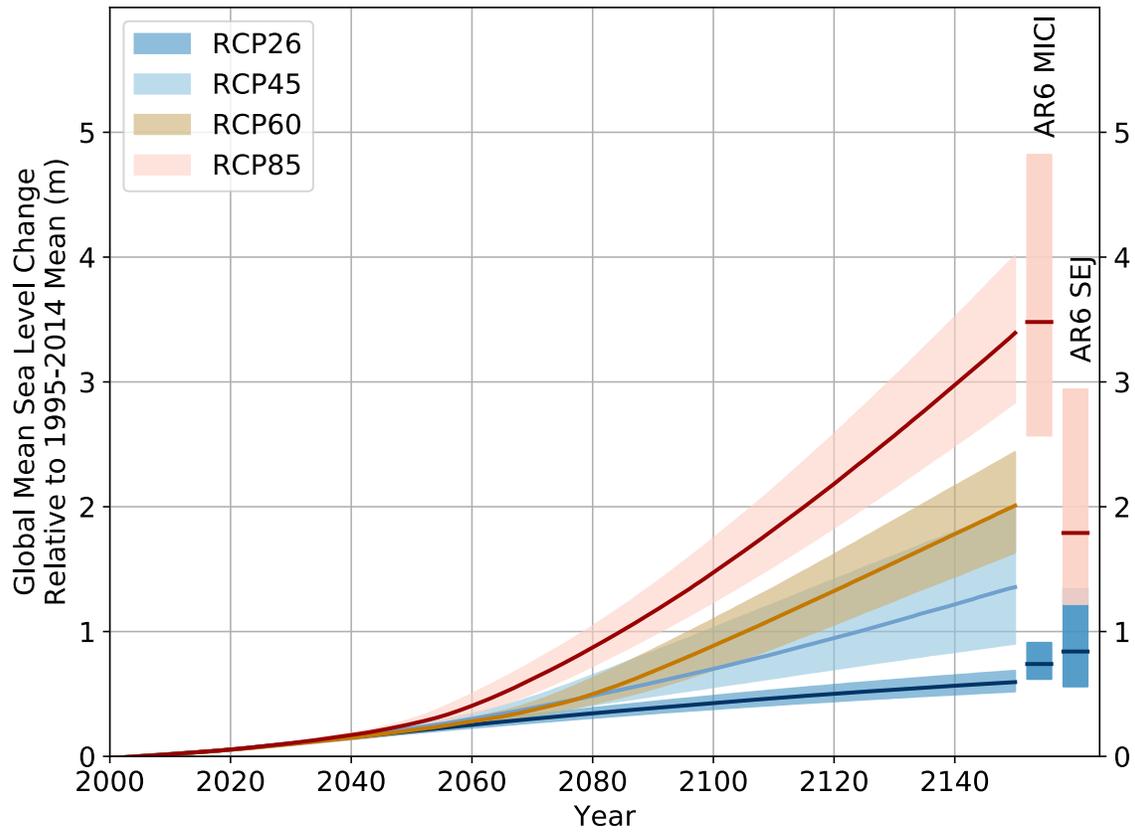

**Figure S2.** Projections of global mean sea level relative to 1995-2014 mean for the four RCP scenarios used in the main text. The vertical boxes on the right denote the Marine Ice Cliff Instability (MICI) and Structured Expert Judgment (SEJ) estimates for 2150 global mean sea level change from the Intergovernmental Panel on Climate Change Sixth Assessment Report (AR6) under RCP2.6 (blue boxes) and RCP8.5 (red boxes) (Fox-Kemper et al., 2021; their Table 9.8). Shaded areas provide the 17-83% (66%) ranges.

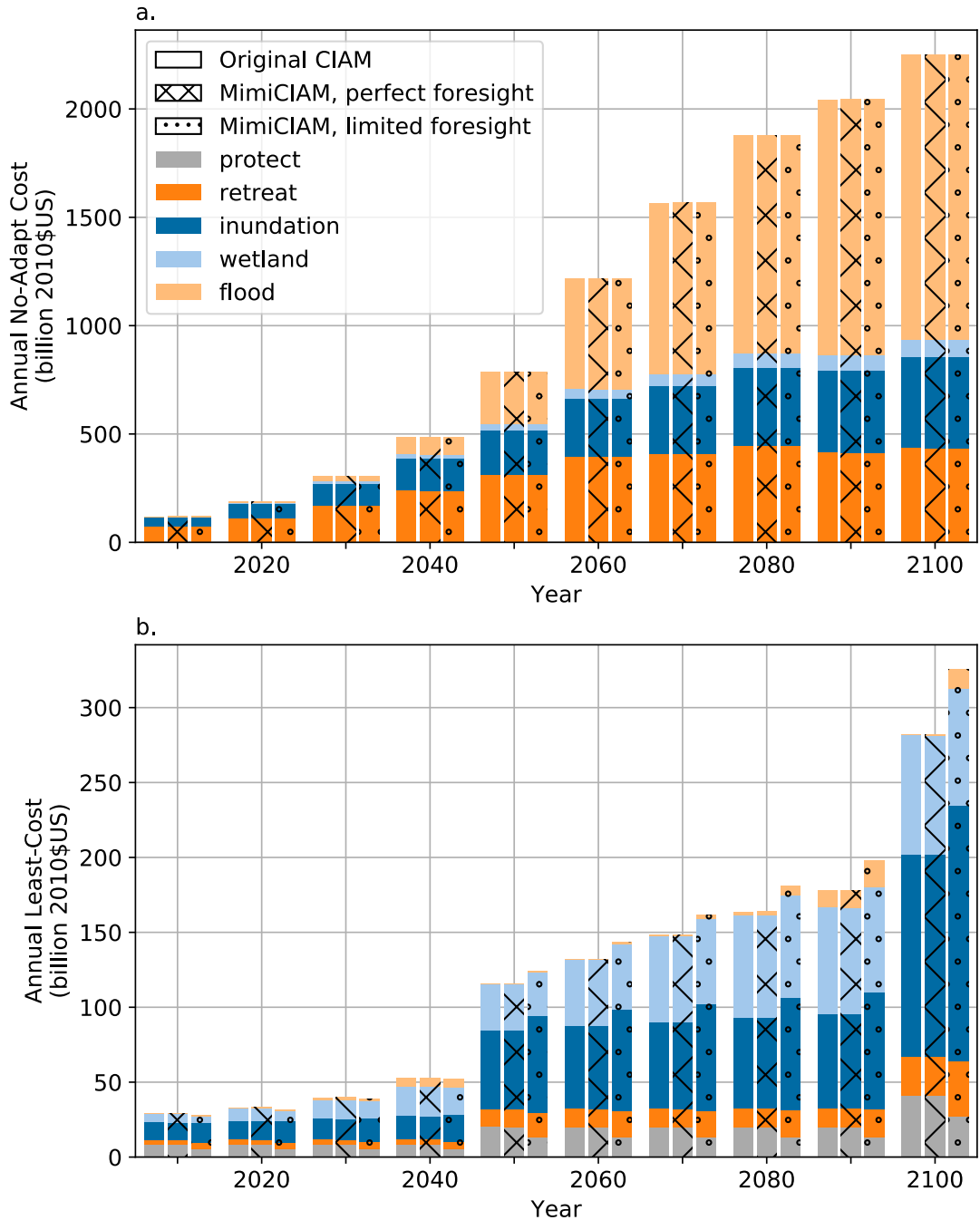

**Figure S3.** Comparison of total global adaptation costs and damages under RCP8.5, following (a) a no-adaptation scenario and (b) least-cost adaptation by minimizing the NPV of total cost over each 40-50-year adaptation period. Model configurations show how MimiCIAM with perfect foresight ("x" hatching) can match Original CIAM (no hatching), and MimiCIAM with limited foresight (stippling) exhibits higher expected costs. To match Original CIAM, the costs from the first time step reference period have not been subtracted from the overall annual costs.

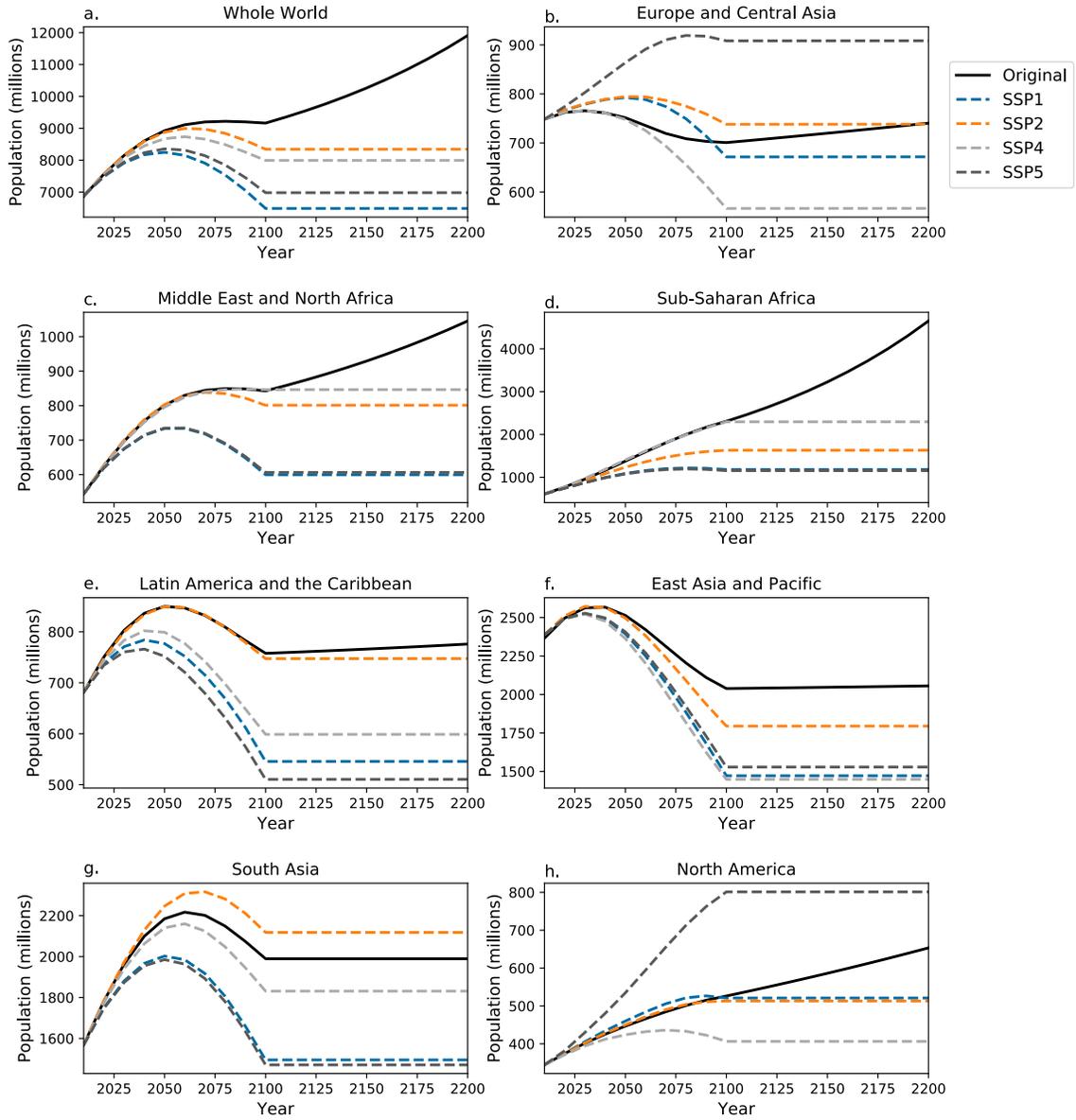

**Figure S4.** Shared Socioeconomic Pathways (SSP) population scenarios, relative to the Original CIAM forcing, aggregated globally and for the seven World Bank regions.

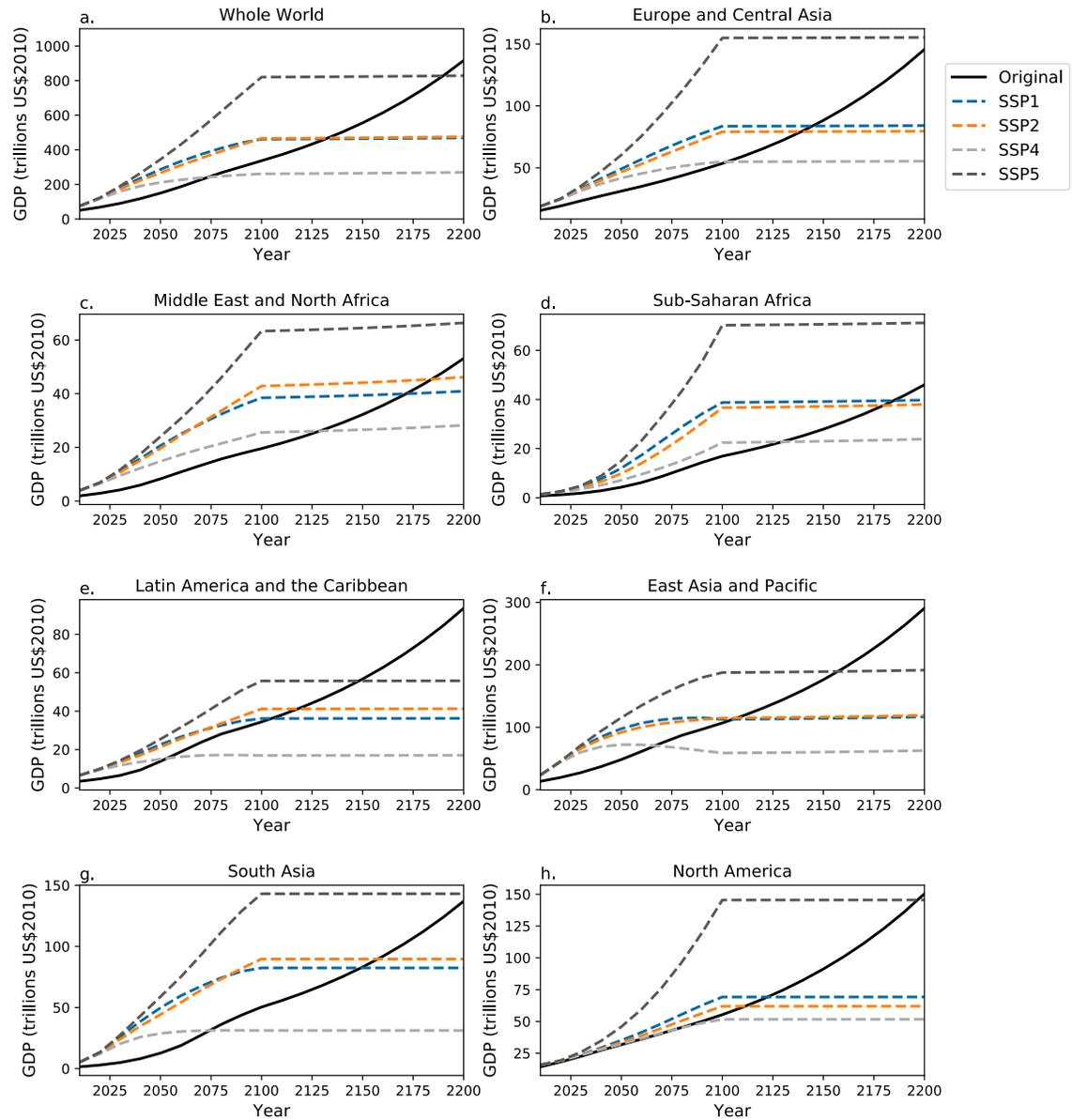

**Figure S5.** Shared Socioeconomic Pathways (SSP) gross domestic product (GDP) scenarios, relative to the Original CIAM forcing, aggregated globally and for the seven World Bank regions.

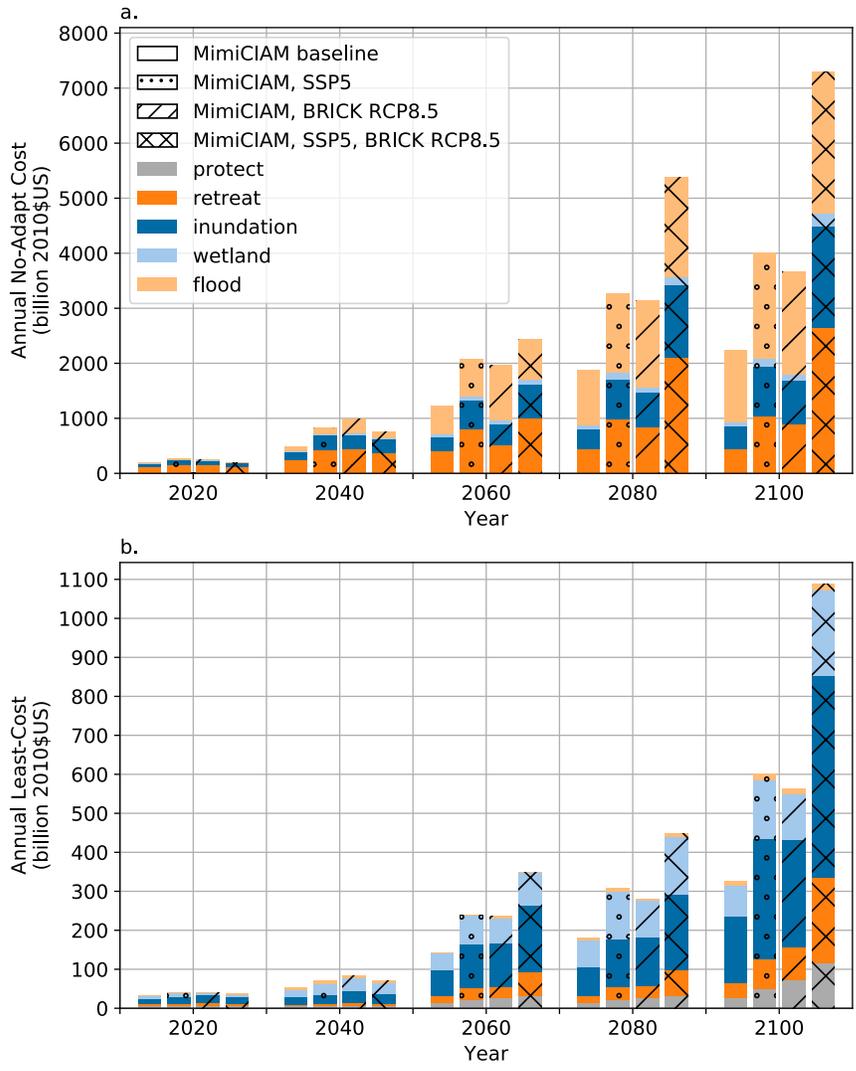

**Figure S6.** Comparison of total global adaptation costs and damages under RCP8.5, following (a) a no-adaptation scenario and (b) implementing least-cost adaptation by minimizing the NPV of total cost over each 40-50-year adaptation period. Model configurations shown are MimiCIAM with limited foresight and all original forcings (no hatching, same as Figure S1), MimiCIAM with population and GDP forcing following SSP5 (stippling), MimiCIAM with sea-level rise forcing from the BRICK model under RCP8.5 (diagonal hatching), and MimiCIAM with both BRICK sea-level rise and SSP5 population and GDP ("x" hatching). To compare against the Original CIAM results, the costs from the first time step reference period have not been subtracted from the overall annual costs.

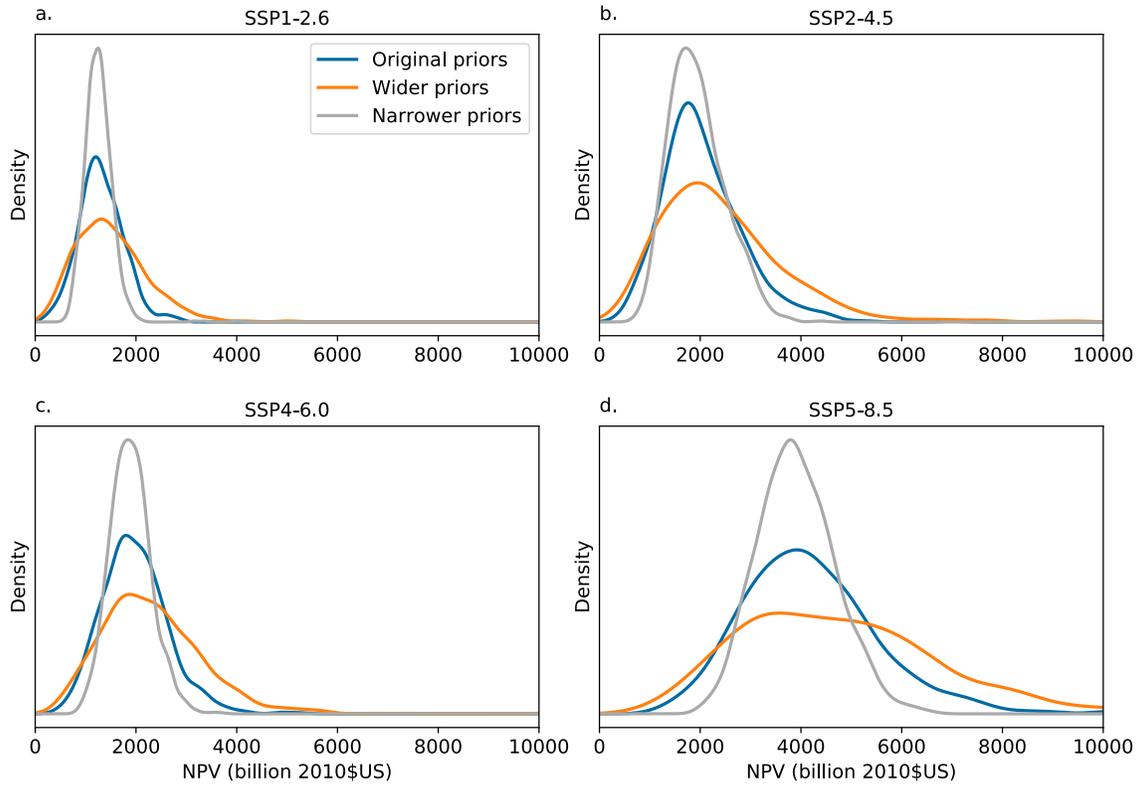

**Figure S7.** Distributions of net present value of total global adaptation costs and damages over the 2010-2150 time period, where the MimiCIAM socioeconomic parameters (Table S1) are sampled from the original prior distributions (blue curves), distributions whose scale parameters are doubled from the original values (orange curves), and distributions whose scale parameters are half of the original values (gray curves). Shown are the four scenarios (a) SSP1-2.6, (b) SSP2-4.5, (c) SSP4-6.0, and (d) SSP5-8.5.

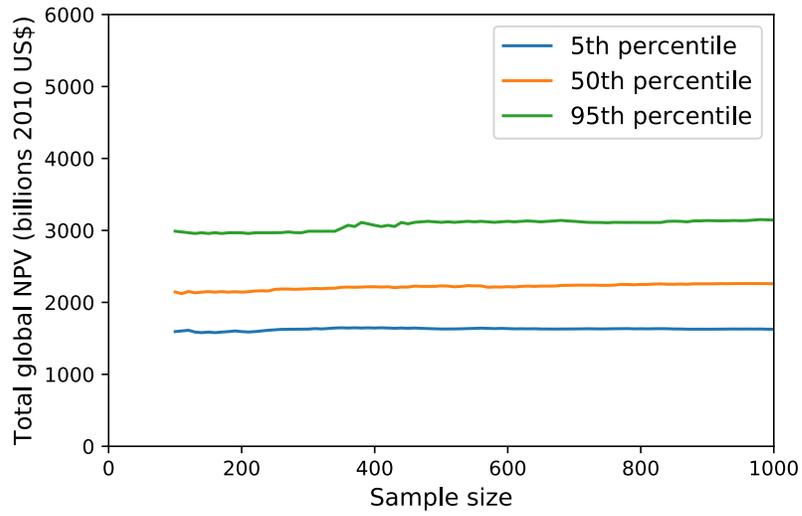

**Figure S8.** Quantiles from smaller subsamples of the Monte Carlo ensemble sampling uncertainty in sea-level rise and socioeconomic parameters, under the SSP2-RCP4.5 scenario.

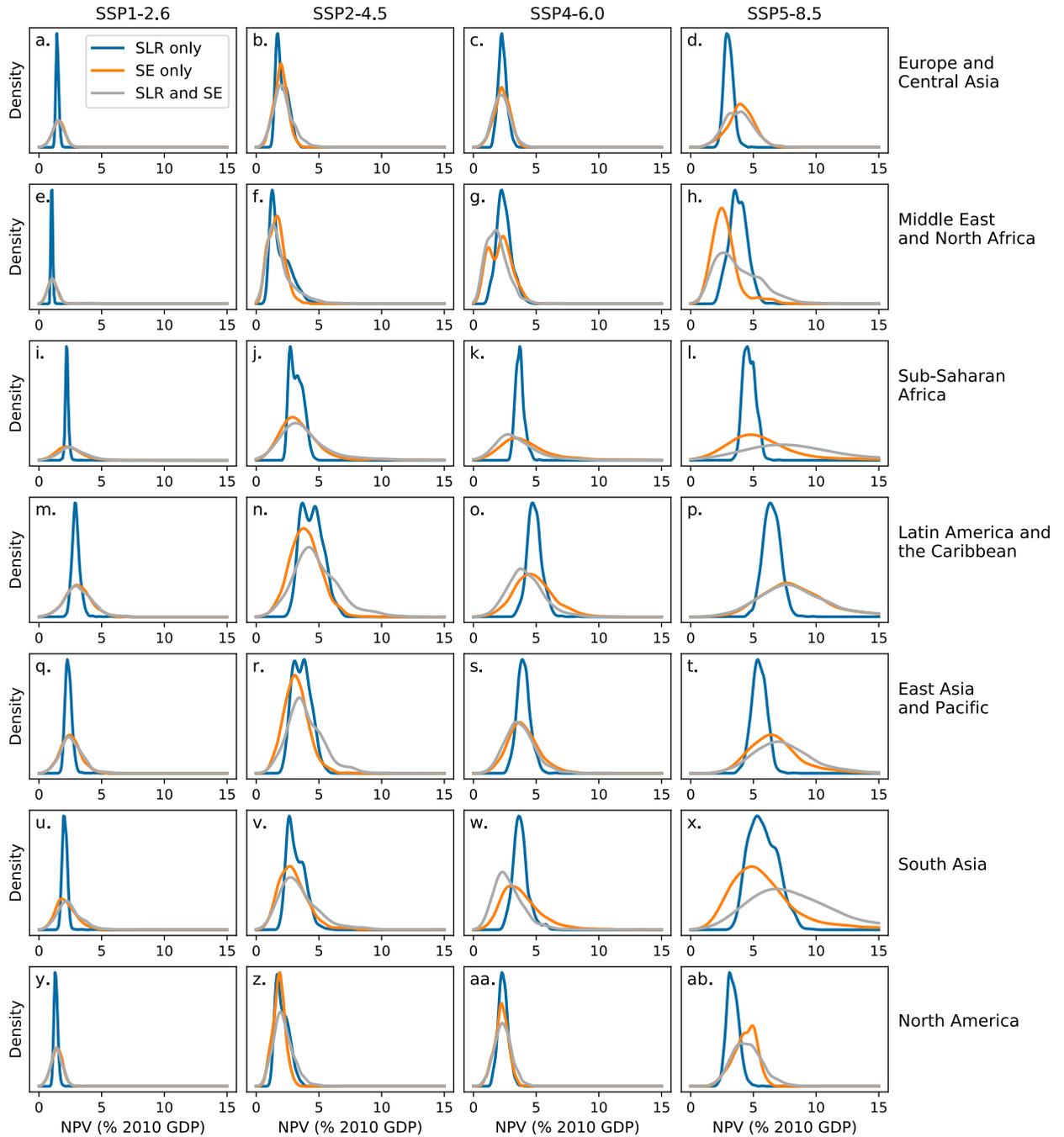

**Figure S9.** Kernel density estimates of least-cost NPV of total adaptation costs over the 2010-2150 time horizon, aggregated for each of the seven World Bank regions: (a-d) Europe and Central Asia, (e-h) Middle East and North Africa, (i-l) Sub-Saharan Africa, (m-p) Latin America and the Caribbean, (q-t) East Asia and the Pacific, (u-x) South Asia, and (y-ab) North America, under pathways SSP1-2.6 (first column), SSP2-4.5 (second column), SSP4-6.0 (third column), and SSP5-8.5 (fourth column). Shown are distributions of adaptation costs accounting for uncertainty in sea-level rise (blue), accounting for uncertainty in socioeconomic parameters (orange), and accounting for uncertainty in both sea-level rise and socioeconomic parameters (gray).

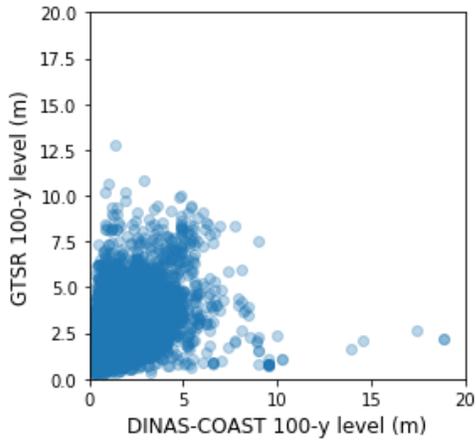

**Figure S10.** Comparison of the 100-year surge exposure levels from the DINAS-COAST and GTSR data sets.

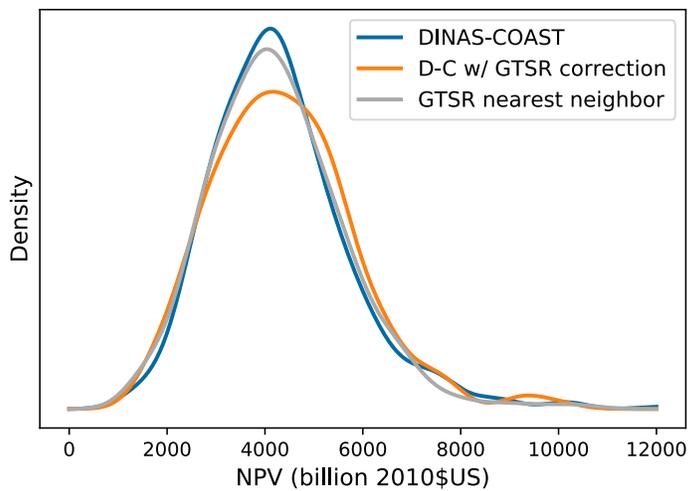

**Figure S11.** Comparison of the net present value of total global adaptation costs and damages over the 2010-2150 time period following SSP5-RCP8.5, using the DINAS-COAST data set ("DC", blue curve), the DINAS-COAST data set bias-corrected using GTSR ("GTSR/DC", orange curve), and the GTSR data set ("GTSR", gray curve).

| Parameter | Units | Central Value | Distribution | Source |
|---|---|---|---|---|
| Relocation cost as fraction of income (movefactor) | Fraction (0-1) | 1 | Normal($\mu=1$, $\sigma=1$), truncated to [0.5, 3] | Anthoff and Tol (2014) and Diaz (2016) |
| Benchmark land value (dvbm) | Million 2010 USD/km$^2$ | 5.376 | Normal($\mu =5.376$, $\sigma =2.688$), truncated to [0, Inf) | FUND; originally Darwin et al. (1995) |
| Population density elasticity of wetland value (wvpdl) | Unitless | 0.47 | Normal($\mu =0.47$, $\sigma =0.12$), , truncated to [0, 1] | Brander et al. (2006) |
| Income elasticity of wetland value (wvel) | Unitless | 1.16 | Normal($\mu =1.16$, $\sigma =0.46$), truncated to [0, Inf) | Brander et al. (2006) |
| Elasticity of value of statistical life (VSL) (vslel) | Unitless | 0.47 | Normal($\mu =0.47$, $\sigma =0.15$) , truncated to [0, Inf) | FUND; Viscusi and Aldy (2003) |
| VSL multiplier on US GDP (vslmult) | Unitless | 200 | Normal($\mu =200$, $\sigma =100$), truncated to [0, Inf) | FUND; originally Cline (1992) |

**Table S1.** MimiCIAM parameters varied in the Monte Carlo uncertainty propagation experiments and their distributions. For convenience, the references cited in the above table are provided below.

| Parameter | Description |
| --- | --- |
| sd_temp | Innovation standard deviation for AR1 residuals for temperature (°C) |
| sd_ocean_heat | Innovation standard deviation for AR1 residuals for ocean heat uptake (10^22 J) |
| sd_glaciers | Innovation standard deviation for AR1 residuals for glacial sea-level contribution (m) |
| sd_greenland | Innovation standard deviation for AR1 residuals for Greenland sea-level contribution |
| sd_antarctic | Innovation standard deviation for AR1 residuals for Antarctic sea-level contribution (m) |
| sd_gmsl | Innovation standard deviation for AR1 residuals for global mean sea level (m) |
| sigma_whitenoise_co2 | White noise standard deviation for CO2 (ppm) |
| rho_temperature | Lag-1 autocorrelation coefficient for temperature (-) |
| rho_ocean_heat | Lag-1 autocorrelation coefficient for ocean heat uptake (-) |
| rho_glaciers | Lag-1 autocorrelation coefficient for glacial sea-level contribution (-) |
| rho_greenland | Lag-1 autocorrelation coefficient for Greenland sea-level contribution (-) |
| rho_antarctic | Lag-1 autocorrelation coefficient for Antarctic sea-level contribution (-) |
| rho_gmsl | Lag-1 autocorrelation coefficient for global mean sea level (-) |
| alpha0_CO2 | Measure of autocorrelation memory for CO2 (-) |
| CO2_0 | Initial CO2 concentration (ppm) |
| N2O_0 | Initial N2O concentration (ppb) |
| temperature_0 | Initial temperature anomaly (°C) |
| ocean_heat_0 | Initial ocean heat uptake (10^22 J) |
| thermal_s0 | Initial condition for thermal expansion sea level contribution (m SLE) |
| greenland_v0 | Initial condition for Greenland ice sheet sea level contribution (m SLE) |
| glaciers_v0 | Initial glacier/ice cap volume (m sea level equivalent (SLE)) |
| glaciers_s0 | Initial condition for glacier/ice cap sea-level contribution (m SLE) |
| antarctic_s0 | Initial condition for Antarctic sea-level contribution (m SLE) |
| Q10 | Respiration sensitivity (-) |
| CO2_fertilization | Carbon fertilization factor (-) |
| CO2_diffusivity | Ocean carbon diffusivity (m/y) |
| heat_diffusivity | Ocean vertical diffusivity (cm^2 s^-1) |
| rf_scale_aerosol | Aerosol radiative forcing scaling factor (-) |
| climate_sensitivity | Equilibrium climate sensitivity to doubling CO2 (°C) |
| thermal_alpha | Global ocean-averaged thermal expansion coefficient (kg m^-3 °C^-1) |
| greenland_a | Equilibrium Greenland ice sheet volume temperature sensitivity (m °C^-1) |
| greenland_b | Equilibrium Greenland ice sheet volume (m SLE) |
| greenland_alpha | Greenland ice sheet response timescale temperature sensitivity (°C^-1 y^-1) |
| greenland_beta | Greenland ice sheet response timescale temperature sensitivity (y^-1) |
| glaciers_beta0 | Initial glacier/ice cap mass balance temperature sensitivity (m y^-1 °C^-1) |
| glaciers_n | Exponent for glacier/ice cap area-volume scaling (-) |
| anto_alpha | Antarctic ocean temperature sensitivity to global temperature (°C °C^-1) |
| anto_beta | Equilibrium Antarctic ocean temperature (°C) |
| antarctic_gamma | Power for relation of Antarctic ice flow speed to water depth (-) |
| antarctic_alpha | Effect of ocean subsurface temperature on ice flux partition parameter (-) |
| antarctic_mu | Parabolic ice surface profile parameter (m^0.5) |
| antarctic_nu | Antarctic runoff and precipitation proportionality constant (m^-0.5 y^-0.5) |
| antarctic_precip0 | Annual Antarctic precipitation for surface temperature 0 °C (m) |
| antarctic_kappa | Coefficient, dependency of Antarctic precipitation on temperature (°C^-1) |
| antarctic_flow0 | Antarctic ice flow at grounding line proportionality constant (m y^-1) |
| antarctic_runoff_height0 | Antarctic runoff line height at 0 °C surface temperature (m) |
| antarctic_c | Antarctic runoff line height temperature sensitivity (m °C^-1) |
| antarctic_bed_height0 | Undisturbed bed height at Antarctic continent center (m) |
| antarctic_slope | Slope of Antarctic ice sheet bed before ice loading (-) |
| antarctic_lambda | Fast Antarctic dynamic disintegration rate (m) |
| antarctic_temp_threshold | Fast Antarctic dynamic disintegration trigger temperature (°C) |

**Table S2.** BRICK model parameters varied in the model calibration and sensitivity analysis.